\pdfoutput=1
\documentclass[sigconf]{acmart}

\copyrightyear{2022} 
\acmYear{2022} 
\setcopyright{licensedusgovmixed}\acmConference[CSLAW '22]{Proceedings of the 2022 Symposium on Computer Science and Law}{November 1--2, 2022}{Washington, DC, USA}
\acmBooktitle{Proceedings of the 2022 Symposium on Computer Science and Law (CSLAW '22), November 1--2, 2022, Washington, DC, USA}
\acmPrice{15.00}
\acmDOI{10.1145/3511265.3550439}
\acmISBN{978-1-4503-9234-1/22/11}

\AtBeginDocument{%
  \providecommand\BibTeX{{%
    \normalfont B\kern-0.5em{\scshape i\kern-0.25em b}\kern-0.8em\TeX}}}

\usepackage{longfbox}
\usepackage{hyperref}
\usepackage{natbib}
\usepackage{amsmath,amsthm}
\usepackage[breakable]{tcolorbox}
\usepackage{physics}
\usepackage{caption}
\usepackage{mathrsfs}
\usepackage{array}

\definecolor{ashgrey}{rgb}{0.9, 0.9, 0.9}

\newcommand{\boxexample}[1]{
\begin{tcolorbox}[breakable]
#1
\end{tcolorbox}
}

\newcommand{\R}{\mathbb{R}}
\newcommand{\X}{\mathbf{X}}

\newcommand{\hmu}{\hat{\mu}}
\newcommand{\E}{\mathbb{E}}
\newcommand{\mc}[1]{\mathcal{#1}}

\newcommand{\bE}{\mathbb{E}}

\newcommand{\bR}{\mathbb{R}}

\newcommand{\kl}[2]{D_{KL}(#1 || #2)}

\newcommand{\ind}{\mathbf{1}}
\newcommand{\reg}{\textsf{regret}}
\newcommand{\rew}{\textsf{reward}}
\newcommand{\Var}{\mathbb{V}}

\renewcommand\footnotetextcopyrightpermission[1]{} % removes footnote with conference information in first column

\begin{document}

\title{Beyond Ads: Sequential Decision-Making Algorithms in Law and Public Policy}

\author{Peter Henderson}
\authornote{Equal contribution.}
\affiliation{%
  \institution{Stanford University}
  \city{Stanford}
  \state{California}
  \country{USA}
}
\author{Ben Chugg}
\authornotemark[1]
\affiliation{%
  \institution{Stanford University}
  \city{Stanford}
  \state{California}
  \country{USA}
}
\author{Brandon Anderson}
\affiliation{%
  \institution{Stanford University}
  \city{Stanford}
  \state{California}
  \country{USA}
}
\author{Daniel E. Ho}
\affiliation{%
  \institution{Stanford University}
  \city{Stanford}
  \state{California}
  \country{USA}
}

\renewcommand{\shortauthors}{Peter Henderson, Ben Chugg, Brandon Anderson, \& Daniel E. Ho}

\begin{abstract}
  We explore the promises and challenges of employing sequential decision-making algorithms -- such as bandits, reinforcement learning, and active learning -- in law and public policy. While such algorithms have well-characterized performance in the private sector (e.g., online advertising),  the tendency to naively apply algorithms motivated by one domain, often online advertisements, can be called the ``advertisement fallacy.'' Our main thesis is that  law and public policy pose distinct methodological challenges that the machine learning community has not yet addressed. Machine learning will need to address these methodological problems to move ``beyond ads.'' Public law, for instance, can pose multiple objectives, necessitate batched and delayed feedback, and require systems to learn rational, causal decision-making policies, each of which presents novel questions at the research frontier.  We discuss a wide range of potential applications of sequential decision-making algorithms in regulation and governance, including public health, environmental protection, tax administration,  occupational safety, and benefits adjudication. We use these examples to highlight research needed to render sequential decision making policy-compliant, adaptable, and  effective in the public sector. We also note the potential risks of such deployments and describe how sequential decision systems can also facilitate the discovery of harms. 
    We hope our work inspires more investigation of sequential decision making in law and public policy, which provide unique challenges for machine learning researchers with potential for significant social benefit. 
\end{abstract}

\begin{CCSXML}
<ccs2012>
   <concept>
       <concept_id>10010405.10010455.10010458</concept_id>
       <concept_desc>Applied computing~Law</concept_desc>
       <concept_significance>500</concept_significance>
       </concept>
   <concept>
       <concept_id>10010405.10010476.10010936.10010938</concept_id>
       <concept_desc>Applied computing~E-government</concept_desc>
       <concept_significance>500</concept_significance>
       </concept>
   <concept>
       <concept_id>10010147.10010257.10010258.10010261.10010272</concept_id>
       <concept_desc>Computing methodologies~Sequential decision making</concept_desc>
       <concept_significance>500</concept_significance>
       </concept>
   <concept>
       <concept_id>10010147.10010257.10010282.10011304</concept_id>
       <concept_desc>Computing methodologies~Active learning settings</concept_desc>
       <concept_significance>500</concept_significance>
       </concept>
   <concept>
       <concept_id>10010147.10010257.10010282.10010283</concept_id>
       <concept_desc>Computing methodologies~Batch learning</concept_desc>
       <concept_significance>500</concept_significance>
       </concept>
 </ccs2012>
\end{CCSXML}

\ccsdesc[500]{Applied computing~Law}
\ccsdesc[500]{Applied computing~E-government}
\ccsdesc[500]{Computing methodologies~Sequential decision making}
\ccsdesc[500]{Computing methodologies~Active learning settings}
\ccsdesc[500]{Computing methodologies~Batch learning}

\keywords{Sequential Decision-making, Reinforcement Learning, Bandits, Active Learning, AI and Society, Law and AI, Responsible AI}

\maketitle

\section{Introduction}
Sequential decision-making (SDM) algorithms have been explored under multiple (overlapping) paradigms, such as bandits, reinforcement learning (RL), and active learning. Together, these have been successfully applied to content recommendation and ad placement \citep{li2010contextual, tang2014ensemble, chen2013combinatorial}, 
clinical trials \citep{durand2018contextual,bastani2020online}, robotics \citep{scholz2014physics}, and control of power systems \citep{reyes2009intelligent,shann2014adaptive}. The core challenge of SDM is carefully balancing the explore-exploit trade-off. 

Resource-constrained problems commonly faced in law and public policy, particularly the public sector, might seem to be a natural but under-explored application for SDM algorithms. Should government agencies focus their resources more on taking advantage of historical information (``exploitation''), or gathering new information (``exploration'')? Too much of the former can result in missing opportunities and trends (e.g. discovering new tax shelters or missing emerging pandemic risk), while too much of the latter results in wasted resources. Sometimes, agencies already use what is effectively an SDM system without a formalization of these trade-offs. Formalization and improved methods may help to address the efficiency, transparency, and accountability of the process.

We present a survey of potential applications and challenges of deploying SDM algorithms in the public sphere.\footnote{We focus primarily on governance in the United States due to the expertise of authors, but most if not all of our examples have parallels in countries and regions globally. 
} The primary audience for this paper is machine learning researchers and engineers. Our central thesis is that off-the-shelf deployment of existing SDM models to the public sector will not work, due to distinct challenges that law and public policy pose. 
Our primary purpose is to highlight these limitations, and provide motivation to the machine learning and computer science communities to work on these frontier problems. To this end, Section~\ref{sec:applications} provides examples of distinct problems faced by various public agencies that \emph{may} be solved, or partially addressed, with further research into SDM algorithms. 

We emphasize that SDM systems, especially those that interact with sensitive data, should be deployed with care. The purpose of this paper is not to suggest SDM algorithms as a one-size-fits-all solution to resource-allocation problems in the public sector. To the contrary, our work spells out the limitations of naively applying methods developed in other domains.
Researchers and practitioners may be tempted to take methods that worked well in online advertising, for example, and apply them without modification to government services. We call this ``the advertisement fallacy.'' 
Algorithm designers are not the only ones susceptible to the advertisement fallacy. By procuring off-the-shelf algorithmic solutions, public agencies have likewise fallen victim to the same error (and been criticized for it~\cite{mulligan2019procurement}). 
We argue that for effective uses in the public sector, machine learning will need to evolve to move beyond ads by solving a number of unique technical challenges. That said, we also illustrate some of the promises of SDMs and we aim to spur more research on specific problems given the potential to increase the efficiency, transparency, and adaptability of government. 
We hope that by discussing specific problems faced by various agencies, and taking the first steps towards formalizing these problems in the language familiar to computer scientists, we can encourage the development of algorithms suited specifically for these challenges.

Our contributions then are: 
\begin{itemize}
    \item A call to shift machine learning research ``beyond ads'' and to law and public policy, where the social gains may be substantial; 
    \item A survey of potential SDM applications in law and public policy, across a wide range of areas, including public health, environmental protection, tax administration, occupational safety, and social welfare;
    \item An enumeration of distinct methodological and technical challenges posed by the public sector that render off-the-shelf application of SDM inappropriate, including multi-objective decision making, batched and delayed feedback, corrupted labels, feedback loops, and rationality requirements; and 
    \item An initial formalization of such challenges to motivate further machine learning research to move beyond ads. 
\end{itemize}

The remainder of the paper is organized as follows. Section~\ref{sec:motivation} discusses the motivations of using SDM algorithms for law and public policy. Section~\ref{sec:background} provides some brief background on the problem formulation. 
Section~\ref{sec:applications} then provides examples of problems faced by various agencies which are potential applications for SDM algorithms and highlights open methodological issues that the machine learning community can address to render such techniques policy-compliant, effective, and more widely applicable. 
We emphasize that while these examples are meant to illustrate interesting research avenues for SDM algorithms, they are \emph{not} meant to indicate that SDMs will certainly solve these problems. Instead, the call is for machine learning researchers to expand the scope of problems they consider when developing algorithms, and to lay some of the foundations for formalizing some of the problems encountered in the public sector. 
Section~\ref{sec:harm} addresses concerns regarding the use of these algorithms, and provides some guidelines to discern when they are appropriate. 

\section{Motivation}
\label{sec:motivation}

As noted, our article highlights frontier challenges that the machine learning research community must address to overcome the advertisement fallacy and address difficult technical challenges faced by the public sector. The examples are meant to illustrate interesting research avenues for SDM algorithms, and not meant to indicate that SDMs will certainly solve these problems. Instead, the call is for ML researchers to expand the scope of problems they consider when developing algorithms, and to lay some of the foundations for formalizing some of the problems encountered in the public sector. These problems demonstrate the broad range of challenges and needs posed by the public sector (e.g., improved data collection). Before we embark on that discussion, however, we hope to briefly discuss several reasons why administrative agencies might \emph{want} to use SDM algorithms in the first place. 

Many public policy processes are already \emph{de facto} SDM algorithms, without formalization. Throughout the subsequent parts of this work, we will provide a number of examples to demonstrate this. We offer several motivations for why formalizing these processes as SDM algorithms may be welfare enhancing, but note that these must be weighed against potential harms (see Section~\ref{sec:harm}).

\textbf{Efficiency.} Improving efficiency of government processes can improve welfare. By formalizing SDM processes, governments can identify bottlenecks in the delivery of important services, identify potential corruption and fraudulent outflows, and improve trust in government through transparency. For example, it has been suggested that investments in information technology at the Internal Revenue Service could lead to 10-to-1 returns on investments by recovering underreported tax payments~\citep{sarin2019shrinking}.

Consider the case of enforcement prioritization at the Environmental Protection Agency (EPA). The EPA must allocate investigative resources to determine who might be in violation of regulations. This naturally requires an explore-exploit trade-off. Without formalization, the EPA may consider allocation via random sampling, or via some other implicit algorithm (perhaps a mix of investigator intuition and other factors). By formalizing the process, the EPA may be able to allocate investigative resources more optimally (e.g., by reducing the amount of random sampling or identifying useful sources of information). 

\textbf{Transparency.} Formalization of processes as SDM algorithms allows for increased transparency, if deployed properly. The data gathered to formally verify SDM models can identify dangerous feedback loops and biases in existing systems. And government oversight organizations (such as the Office of Inspector General) and policy makers, can leverage this data to assess and adjust government services.

\textbf{Adaptive Policymaking.} It is often not clear whether to deploy an algorithm in a new context. Algorithmic impact assessments have been suggested as a solution to the problem of how and if algorithms should be deployed, aiming to foster reflection about the risks of adoption~\citep{metcalf2021algorithmic}. They have been adopted by several governmental organizations, such as the EU \citep{kaminski2020algorithmic} and Canada \citep{mckelvey2019artificial}. Such reflection may rule out the deployment of algorithms with inherent harm. 
However, as noted by \citet{moss2020governing}, there is disagreement concerning when algorithmic impact assessments should be conducted. Some advocate for \emph{ex ante} assessments, i.e., occurring before the adoption of the proposed system, while others stress the importance of \emph{ex post} reviews,\footnote{A recent example is Facebook's study of its role in Myanmar \citep{warofka2018independent}.} which take place after the system has been implemented. The former emphasize that systems adopted without heavy scrutiny could cause avoidable harm, while the latter point out that that insufficient information exists at the time of the adoption decision. \emph{Ex ante} assessments, in other words, require predicting the consequences of algorithmic decision-making. Costs and benefits are not precisely known. 

SDM algorithms can help narrow the divide between the \emph{ex post} and \emph{ex ante} camps. Because they are sequentially implemented, they enable real-time tailoring to adapt their behavior. For instance, we might increase the amount of random exploration conducted by the algorithm, or reduce the weight of previous time period which is unrepresentative of the larger population. Much like in the case of adaptive clinical trials -- where sequential assignment enables researchers to limit the number of patients assigned to ineffective treatment arms -- SDMs may enable ongoing assessments of an algorithmic impact. The stochastic component of an ``explore'' decision potentially enables researchers to ascertain the impact of an intervention, thereby curing the information deficit at the time of adoption. This epistemic benefit to SDMs may also enable more responsible adaptation, scaling, and discontinuation of algorithms in the public sector.

We also note that error-correction may be easier in formalized automated decision-making systems than in \emph{ad hoc} decision-making systems. The former yield to statistical and mathematical insights, enabling us -- if the system is implemented with transparency in mind -- to determine why certain decisions were made, and to correct for unintended outcomes.
Recent work, for instance, has developed algorithms to explicitly trade-off between statistical analysis with reward in adaptive experiments~\citep{nogas2021algorithms}. 
This is in contrast to ad hoc decision-making systems, which often elevate the status quo and implicit reasoning, rendering them impermeable to criticism. 
However, all of the above motivations are moot if technical challenges are not solved and harms are not addressed. The subsequent sections provide an agenda for tackling this.

\section{Problem Formulation}

\begin{table*}[t]
\centering
\begin{tabular}{
p{4cm}|p{6cm}|p{6cm}}
 \toprule
 \textbf{Framework} & \textbf{Goal} & \textbf{Example Use Case}
 \\
 \midrule
 Bandit & Minimize regret by selecting rewarding arms & Allocate inspections to buildings most in need of repair\\
 \hline
 Active learning & Minimize generalization error by selecting examples which most improve the model & Construct, at lowest cost, a maximally accurate model of building inspection need \\
 \hline
 Reinforcement learning ($0<\gamma<1$) &
 Maximize future \emph{discounted} reward by constructing explicit policy  
  & Learn an optimal maintenance schedule by minimizing future maintenance costs \\
\bottomrule
\end{tabular}
\caption{A stylized breakdown of the three frameworks and their goals, along with an example use case from infrastructure management. }
\label{tab:lenses}
\end{table*}

\label{sec:background}
The interface between public policy and SDM encompasses a wide variety of both tasks and methodological approaches. In order to ease the navigation of disparate settings and paradigms, we use this section to provide a general problem formulation and vocabulary. We also consider a basic hypothetical example and then modify it to demonstrate how and under what circumstances each of the three major SDM paradigms -- multi-armed bandits, reinforcement learning, and active learning -- might make for a sensible approach. 

As the name implies, SDM algorithms are sequential, with decisions taking place across a time series. We constrain discussion to those algorithms that use discrete \emph{timesteps}  $t=1,\dots,T$. These timesteps can correspond to anything from tax years to court cases. At each time $t$, the \emph{agent} receives a set of $N_t$ observations, the $i$-th observation having features (sometimes referred to as \emph{context}) $x_{it}\in \R^d$. Denote the collection of all observations at time $t$ as $\X_t=(x_{it})_{i=1}^{N_t}$. Observation $i$ has a hidden label or \emph{reward} $r_{it}$. In the regulatory setting for example, reward can be expressed as a quantitative measure of compliance, only observable with a directed \emph{action}: the monetary value of a tax adjustment is obtained by performing an audit, or the number of food safety violations at a restaurant by inspection. A \emph{policy} decides which set of $m_t$ actions to take, with actions typically requiring selecting $m_t$ observations. These actions then yield rewards. Typically, $m_t\ll N_t$ though the set of observations and selection budget may change size over time ($N_t\neq N_{t+1}$ and $m_t\neq m_{t+1}$).

To clarify the presentation of the three main approaches to sequential decision-making under consideration -- active learning, bandits, and reinforcement learning -- we introduce the running example of civil infrastructure management. Consider a city or other agency which must plan the maintenance and rehabilitation of public infrastructure.  
Each time period, say yearly in this case, the agency has a limited budget with which to inspect and perform maintenance on the global pool of projects. They might have prior information pertaining to each project, including the results of any previous inspections, and perhaps physical models of resiliency and decay. This problem is not hypothetical, and has been subject to various approaches (see \citep{kabir2014review} for an overview), many from stochastic control, and some from RL \citep{madanat1993optimal,andriotis2021deep}. The goal of the SDM algorithm is to optimize resource allocation by recommending an optimal set of infrastructure projects to either visit or  rehabilitate (depending on the specifics of the problem). As we will see, different frameworks will be more or less suitable depending on the problem and objective.

\subsection{Bandits}
A natural goal for the agency is to allocate infrastructure inspections to those projects that are most in need of them. 
Perhaps the most natural framework in this case is the multi-armed bandit formulation. Here, there are $N$ ``arms'' to choose from, each associated with its own reward distribution. At each time step, we choose an arm (or set of arms given the \emph{batched} setting, see e.g.,~\citet{perchet2016batched}).
In our example, arms might be specific projects, or they could be specific classes of projects (e.g., bridges, pipes, office buildings, apartments, hospitals, etc.).  

If incorporating prior knowledge -- \emph{observations} -- of each project (or class of projects), the problem can be cast as a batched, \emph{contextual bandit}, in which the observations comprise the context.\footnote{Alternatively, observations might be grouped along shared features, in which case we can treat the problem as a contextual bandit in which arms are groups of observations~\citep{soemers2018adapting}.} 

If we additionally assume that rewards can be approximated as a functional form of observations, we can interpret the problem naturally as a structured bandit~\citep{mersereau2009structured, auer2002using}. In the bandit setting, one of potentially multiple objectives (see Section~\ref{sec:multi-obj}) is maximizing cumulative reward over time (equivalently, minimizing cumulative regret). Cumulative reward and cumulative regret at time $t$ are defined as 
\begin{equation}
    \label{eq:reward}
    \rew(t) \equiv \sum_{\tau\leq t} \E\bigg[\sum_{i\in S_\tau} r_{i\tau}\bigg], 
\end{equation}
and 
\begin{equation}
\label{eq:regret}
\reg(t) \equiv \sum_{\tau\leq t} \E\bigg[\sum_{j\in S^*_\tau}r_{j\tau} - \sum_{i\in S_\tau} r_{i\tau}\bigg]
\end{equation}
where $S_t$ is the (possibly singleton) set of actions we take at time $t$, and $S^*_t$ is set of actions with the highest total reward. The expectation is taken over any random choices made by the algorithm, as well as over stochastic rewards. In the infrastructure example, reward might be whether, or to what extent, the project requires maintenance. In this case, regret measures how many sites were visited that don't require timely attention. 

\subsection{Active Learning}
Another possible objective in our example is to train a fixed model to predict maintenance requirements while meeting a strict inspection budget. This falls more naturally into an active learning paradigm \citep{settles2009active}. 
The goal of active learning is to maximize accuracy of the underlying model while making as few queries as possible. Given a pool of unlabeled observations ($\X_t$) (infrastructure projects) the policy would decide which label ($r_{it}$) to reveal (which inspection to perform) to maximize the model's performance across the pool. Thus the reward is the reduction in generalization error from selecting a given arm. In some active learning formulations (like continuous active learning), the goal can instead be framed as successfully identifying all the labels of a given type~\citep{cormack2015autonomy}. In this case $r_{it} = 1$ if $y_{it}$ is of the desired type.

\subsection{Reinforcement Learning}
If the agency would like to develop an explicit future maintenance schedule then reinforcement learning is an appropriate paradigm. RL enables us to develop a policy which minimizes future costs,\footnote{We note that contextual bandits can be considered a form of reinforcement learning, but for simplicity we distinguish between the two here.} and is particularly valuable when there is an underlying model for how various ``states'' relate to one another. 

We define the reinforcement learning problem as an infinite-horizon discounted Markov Decision Process (MDP)~\citep{bellman1957markovian,Puterman94}. An MDP is defined as $\mc{M} = \langle \mc{S}, \mc{A}, \mc{R}, \mc{T}, \beta, \gamma \rangle$ where $\mc{S}$ is the state space, $\mc{A}$ is the action space, $\mc{R}:\mc{S} \times \mc{A} \mapsto \bR$ is the reward function, $\mc{T}:\mc{S} \times \mc{A} \mapsto \Delta(\mc{S})$ is the transition function, a kernel mapping state-action pairs to a probability distribution over $\mc{S}$, $\beta \in \Delta(\mc{S})$ is the initial state distribution, and $\gamma \in [0,1)$ is the discount factor. In the applications we study, more often than not, multiple actions will need to be taken per time-step and the transition function will not necessarily be known \emph{a priori}. The state space will be constructed from $\X_t$, the set of arms available for selection. Furthermore, in many cases the true problem formulation may be better suited for the partially observable MDP framework~\citep{cassandra1998survey}, as the state space is unlikely to be fully observable. Unlike the bandit setting which maximizes myopic reward, the RL setting maximizes the expected discounted future return at each timestep, \begin{equation}
\label{eq:return}
    V^\pi(s) = \bE\bigg[\sum_{t=0}^\infty \gamma^{t} \mc{R}(s_t, a_t) | s_0 = s\bigg].
\end{equation}
\citet{madanat1993optimal} casts the infrastructure management problem as a MDP, where the states correspond to a project's condition, and transition probabilities capture deterioration over time and the effect of performing various kinds of maintenance. \citet{andriotis2021deep} use RL to optimize a reward function that would minimize total discounted future costs, including cost of providing maintenance versus costs from deterioration.

\section{Applications \& Obstacles}
\label{sec:applications}
Here we discuss various applications of SDM in law and public policy. Along with the examples, we highlight specific problems to be solved in each area to ensure that these algorithms are policy compliant, effective, and reliable. For each example, we also give a possible formulation of the problem as an SDM algorithm.

We reiterate that we are choosing this set of problems because they demonstrate the broad range of challenges posed by the public sector and what might need to be improved to prepare the public sector for such uses of machine learning (e.g., better data collection), not because they readily lend themselves to algorithmic solutions. 

This section should not be read as a comprehensive list of opportunities and challenges of applied ML. 
Much prior work has focused on challenges of deploying ML systems in various areas such as health, education, and the sciences (e.g., \citet{ghassemi2020review,karpatne2018machine,beam2018big}). Privacy, fairness, and explainability, for instance, are important and  well-documented concerns generally~\citep{papernot2016towards,mehrabi2021survey,amarasinghe2021explainable}.
We focus on challenges that are of particular prominence for SDM in law and public policy, and especially the public sector, aside from these well-documented areas of research. For each challenge we identify potential use-cases in real-world settings. Table~\ref{tab:laws} provides examples of distinct technical challenges that arise from legal and policy constraints.

\begin{table*}[!htbp]
\begin{center}
\resizebox{\textwidth}{!}{
\begin{tabular}{ | p{1in} | p{1.9in}| p{1.25in}|  p{2.5in} | } \hline
Technical Challenge & Public Policy Example & Example Organization(s) & Example Legal and Policy Constraint(s) \\
  \hline 
 \textbf{Multi-objective decision-making} 

 & \underline{eDiscovery}: maximizing the retrieval of responsive documents in discovery, while estimating the prevalence of responsive documents & U.S. Courts & Federal Rules of Civil Procedure Rule 26(g) (creating ``reasonable inquiry'' standard for the production of documents during discovery, often taken to mean a high recall rate for retrieval models~\citep{grossman2014comments,henderson2022discovery}). \\
 & \underline{Identifying Tax Misreporting}: maximizing discovery of tax evasion while estimating population prevalence of income misreporting  & Internal Revenue Service & Internal Revenue Code \S~7602; Internal Revenue Manual 1.2.1.5.10 (stating policy around use of ``statistics indicating the probability of substantial error''); Improper Payments Elimination and Recovery Act (requiring a ``statistically valid estimate'' of improper payments); Office of Management and Budget, Requirements for Payment Integrity Improvement (spelling out methodological considerations for estimation) \\
  \hline
  \textbf{Batched and delayed feedback} & \underline{Anti-corruption}: anticorruption audits can be risk-selected, but feedback can be non-randomly delayed by years & Brazilian Controladoria Geral da União (CGU) & Law \textnumero  10683 of 28 May 2003 (Brazilian law creating the CGU which conducts anticorruption audits)\\ 
  \hline
  {\bf Distribution shifts in the small data regime} & \underline{Public health pandemic response}: incorporating exogenous and endogenous changes in disease rates into predictions & Local Public Health Departments, U.S. Centers for Disease Control and Prevention & Executive Order 13,994 (requiring agencies to investigate new data collection mechanisms and data-driven responses to public health threats);  Executive Order 13,996 (establishing a board to allocate resources for testing, while prioritizing at-risk communities); \citep{krass2021us} (discussing additional legal constraints).\\
   & \underline{Food Safety}: maximizing the discovery of food safety violations while estimating the prevalence of non-compliance in import inspections & U.S. Food and Drug Administration & Food Safety and Modernization Act \S~201 (Regulations regarding the methods for "Targeting Of Inspection Resources For Domestic Facilities, Foreign Facilities, and Ports Of Entry") \\
  \hline 
  {\bf Learning with and identifying corrupted labels} & \underline{Occupational safety}: Allocating inspectors to visit facilities to ensure occupational health and safety requirements are met  & Occupational Safety and Health Administration& Occupational Health and Safety Act, 29 U.S.C. \S\S~657, 667 (providing for inspections and requiring that state enforcement must be at least as effective as federal enforcement); Administrative Procedure Act, 5 U.S.C \S~706 (providing that a reviewing court shall set aside agency action that is arbitrary and capricious); \cite{ho2017does, morantz2009has} (discussing inconsistencies in inspection process and state enforcement)\\
  \hline
  {\bf Feedback Loops} & \underline{Environmental compliance}: Determine which facilities to inspect, and the method of inspection, to ensure environmental regulations are upheld & U.S. Environmental Protection Agency & Clean Water Act \S~308 (providing for inspection authority); National Compliance Initiative (aiming for reduction in overall noncompliance rate); \cite{gao_epa} (documenting data gaps to monitor overall levels of compliance) \\
  \hline 
  {\bf Rational and Causal Decision-Making} & \underline{Disability adjudication}: Determining the subset of claims likely to be approved so that they can be flagged for speedy review & Social Security Administration & 20 CFR \S\S~404.1619, 416.1019 (providing for quick disability determination if there is a ``high degree of probability that the individual is disabled''); Motor Vehicle Mfrs. Ass'n v. State Farm Mutual Automobile Ins. Co., 463 U.S. 29 (1983) (articulating standard for whether agency has considered the relevant factors and reached a reasoned decision)\\
\hline
  {\bf Validation} & \underline{Wildfire prevention targeting}: Identify forest regions to actively manage, including conducting prescribed burns, to reduce wildfire impacts. & U.S. Department of the Interior & John D. Dingell, Jr. Conservation, Management, and Recreation Act (2019); Executive Order 13,855: Promoting Active Management of America's Forests, Rangelands, and Other Federal Lands To Improve Conditions and Reduce Wildfire Risk\\
  \hline
\end{tabular}}
\caption{A subset of the examples discussed throughout the paper, listed alongside (one of) their core technical challenges and any legal requirements which place constraints on the problem. We note that other legal constraints also apply to all of the presented algorithms, including privacy and non-discrimination standards. This table is meant as a non-exhaustive illustration of the distinct challenges posed by law and public policy.}
\label{tab:laws}
\end{center}
\end{table*}

\subsection{Multi-Objective Decision-Making}
\label{sec:multi-obj}

Agencies can rarely afford to be single-minded. 
The Environmental Protection Agency (EPA), for instance, might wish to identify the highest polluters and penalize them. But they may also need to accurately estimate the overall level of noncompliance by regulated facilities~\citep{gao_epa}. Statutory obligations may specifically require that agencies estimate population quantities~\citep[e.g.,][]{omb2018requirements}.
In general, government agencies must both target risk and estimate total noncompliance  to effectively guide policy. 

Such an objective of \emph{population estimation} --- which have conventionally been less of a focus for the SDM literature --- may need to integrate methods from survey sampling \citep{lohr2019sampling}. Indeed, methods such as Horvitz-Thompson sampling \citep{horvitz1952generalization} have been recently adapted to multi-armed and structured bandit settings to achieve unbiased population estimates \citep{henderson2022maximize, chugg2021reconciling}. 
There has also been work on understanding population estimation in bandits without considering reward as an additional objective. In multi-armed bandits, for instance, it is understood that the sample mean is not an unbiased estimate of each arm because the data is collected adaptively \citep{nie2018adaptively,shin2021bias,shin2019sample}. Leveraging such insights in settings where more than one objective is being considered would be valuable, especially with the aim of analyzing trade-offs.

Population estimation is related to so-called \emph{active exploration}, which has been studied both in bandits \citep{carpentier2011upper,antos2010active} and in reinforcement learning and MDPs \citep{shyam2019model,tarbouriech2019active, thrun1991active}. Active exploration is focused on efficiently exploring an unknown space to make accurate predictions about unlabelled observations.  
\citet{erraqabi2017trading} study how to trade-off active exploration with reward maximization in the multi-armed bandit setting. Developing algorithms to trade off these two objectives in other settings remains an open problem. 

Agencies may have multiple secondary objectives. For instance, educators wish both to determine the effectiveness of various educational policies, but also maximize students' learning in the meantime~\citep{liu2014trading}. That is, spending resources on exploring new policies might disrupt the classroom and hamper learning. Attempts to handle generic multi-objective problems often involve extending the reward scalar to a vector~\citep{drugan2013designing}, and searching for Pareto optimal solutions. While such approaches are useful for generic problems, they discard the possibility of taking advantage of specific objectives that agencies commonly face.

\boxexample{
\begin{example}
[\textbf{Optimizing tax audits at the IRS}] Every year, the Internal Revenue Service (IRS) audits between 0.5\% and 1.1\% of individual taxpayer returns, out of a potential pool of hundreds of millions~\citep{sarin2019shrinking}.
In addition to maximizing revenue, they must generate reliable population estimates (the ``tax gap''). Moreover, these estimates are often subject to accuracy standards set by the Office of Management and Budget (OMB) \cite{omb2018requirements}.
\end{example}

\textsc{Formulation.} This was recently modeled as a structured bandit problem \citep{henderson2022maximize}.  Each year the IRS selects a batch of taxpayers to audit from the population based on their featurized tax returns $x_{it}$. The reward received, $r_{it}$, is the difference between the taxpayer reported amount of taxes owed and the true adjusted amount after audit. The reward is related to taxpayer's attributes $x_{it}$ via some unknown function $f$ and parameter $\theta_t$: $r_{it}=f(x_{it},\theta_t)$. 
One objective is to maximize reward over time, i.e., Equation~\eqref{eq:reward}.  Another  is to estimate the average adjustment each year, $\mu_t=\sum_i r_{it}$. 

     There are several ways in which we might formalize the objective. One is to minimize a convex combination of these objectives, e.g., 
     \[\min_{S_1,\dots,S_t} \bigg(\reg(t) + \alpha \sum_\tau|\hmu_\tau-\mu_\tau|\bigg),\]
     where $\alpha$ is some normalizing parameter which weights the importance of accurate population estimation against regret.  A separate approach is to treat it as a constrained optimization problem in which we maximize reward subject to limits on the expected percent difference and/or the variance, in adherence with OMB guidelines for precision of estimates: 
     \begin{align*}
         \max_{S_1,\dots,S_t} &\quad \rew(t) \\ 
         \text{s.t.} & 
         \quad  |\E[\hmu_\tau]-\mu_\tau|/\mu_\tau\le \beta, \\ 
         & \quad \Var(\hmu_\tau) \le \kappa \quad\text{for all } \tau\leq t,
     \end{align*}
     for some $\beta,\kappa>0$. 
}

\begin{tcolorbox}[breakable]
\begin{example}[\textbf{Technology Assisted Review during civil litigation}]
    eDiscovery, short for electronic discovery, entails identifying relevant documents during the discovery process in legal proceedings. Since the body of potential evidence is often overwhelming, current state-of-the-art eDiscovery mechanisms use active learning to identify relevant documents~\citep{chhatwal2017empirical,cormack2015autonomy,henderson2022discovery}.
    The use of technology assisted review is especially common in antitrust litigation by the Department of Justice and Federal Trade Commission.
    We also note that identifying documents for FOIA requests uses a similar process in some cases.
\end{example}
    
     \textsc{Formulation.} The policy sees a set of documents (arms) $\X_t$, with $x_{it}$ consisting of standard natural language processing features (e.g., TF-IDF vectors \citep{jurafsky2014speech}). The policy selects a set of documents for lawyers to review and receives a reward if the document was responsive ($r_{it}=1$) or non-responsive ($r_{it}=0$). This is repeated across multiple rounds until there is confidence that all responsive documents have been identified. Here we're  trying to minimize the number of non-responsive documents, i.e., minimizing $\sum_i r_{it} - \sum_{i\in S_t} r_i$, while simultaneously estimating $\mu_t$. In current implementations,  
     a random sampling mechanism is often employed to estimate $\mu_t$~\citep{li2020stop}. Combining these two processes into one multi-objective system might help to reduce labeling costs. 
\end{tcolorbox}

\subsection{Batched and Delayed Feedback}
\label{sec:batching}
In contrast to online advertisements, it is rare to witness the result of each action before having to make subsequent decisions in law and policy. 
Instead, decision-making must be batched and could potentially be based on outdated information. This leads to ask whether we can incorporate such features into SDM algorithms. %
There has been work on both batched and delayed feedback in the bandit setting~\citep{huang2016linear,lansdell2019rarely,gao2019batched,esfandiari2021regret}, as well as the active learning setting~\citep{guo2007discriminative, citovsky2021batch}. In multi-armed bandits in particular, concept drift is sometimes modelled as discrete, abrupt changes in the mean reward of the arms \citep{mellor2013thompson,alami2017memory}, sometimes called \emph{switching bandits}. 

Unfortunately, such algorithms typically make unrealistic assumptions, such as knowing an \emph{a priori} bound on the amount of drift or the total number of time steps in order to optimally partition the actions in pre-defined time blocks~\citep{russac2019weighted,cavenaghi2021non,zhao2020simple}. 
The existing work on batched feedback in bandits suffers similar drawbacks~\citep{gao2019batched,perchet2016batched}.

In law and policy settings, delays can be stochastic, biased, and dependent on arm features, relationships which aren't considered in prior work. For example, an audit of a small facility will likely be finished more quickly than a large one. Variance in delays can also be heteroskedastic: more complicated audits could have a wider range of delays.
Moreover, concept drift and batched actions have not been studied in conjunction with one another, whereas they often appear together in policy settings.

\boxexample{
\begin{example}
[\textbf{Auditing local government corruption at the CGU}] In 2003, the government of Brazil introduced the Controladoria Geral da Uni\~{a}o (CGU), an autonomous federal agency to conduct anti-corruption audits of local governments~\citep{ferraz2008exposing}. While the CGU audited local governments at random, \citet{ash2021machine} demonstrated that selection informed by tree-based classifier would catch nearly twice as much corruption. 
Auditing, however, will change the subsequent behavior of departments and can take years to complete. Therefore, all audits for the next fiscal year have to be chosen before receiving any feedback and will deal with outdated information. Time to completion for an audit will also likely correlate with the local government's size and composition. The heterogeneity of the delay schedule might affect algorithmic biases if not corrected for appropriately. 
\end{example}

\textsc{Formulation.} Each year, the policy can select a set of local governments to audit given features consisting of tax revenue, size of government, history of previous corruption, etc.
The reward consists of successfully identified sources of corruption, but may take several years (time steps) to return and update the policy. In particular, when an audit is conducted it takes time $d_{it}\sim \mathcal{D}_t$ to receive reward $r_{it}$, where $\mathcal{D}$ is some unknown delay distribution. While previous work assumes a single delay distribution, or one that varies with time (e.g., \citet{bistritz2019exp3} and \citet{ vernade2020linear}), here the distribution is most likely affected by the government itself. We can model this as a family of distributions $\mathscr{D}=\{\mathcal{D}_\theta\}$ parameterized by latent variables $\theta$. In addition to identifying corruption, one should learn the parameters $\theta$ which give rise to delays and the model must account for censoring by marginalizing over audits that have not yet cleared. To complicate things further, we must select all audits in year $t$ as a batch, without receiving feedback from any of them. We might also expect partial feedback of the audits, in which we receive some signal each year until we receive the true result~\citep{grover2018best}. 
}

\subsection{Distribution Shifts in the Small Data Regime}
The distributions of both policy-relevant features and their relationship with reward are rarely fixed in reality—behaviors adapt, policies change, exogenous economic shocks occur, etc.  This \emph{concept drift} emphasizes the exploration aspect of SDM, requiring careful selection just to keep up.  Algorithms in public policy settings are doubly hampered as the amount of available samples is often small, potentially leaving little budget for reliably rewarding selections.  

Recent work in both the bandit and active learning spaces focuses on either detecting concept drift or designing algorithms which are robust to it~\citep{cavenaghi2021non, soemers2018adapting, krawczyk2019adaptive}. 
Such work, however, develops algorithms that are purely reactive. That is, they do not anticipate concept drift. Meanwhile, in law and public policy settings, there is often specific domain information which can act as an \emph{a priori} indicator of concept drift (e.g., a statutory revision). Incorporating this structured drift information into existing models could improve resilience and efficiency, particularly in small-data regimes. More research is needed to create reliable methods for incorporating such structural priors or external sources of information into SDM policies. Moreover, it is an open question as to which algorithms retain performance and fairness guarantees under such large discontinuous sources of drift.

The scarcity of data in such conditions makes deploying reinforcement learning algorithms more difficult. RL algorithms often require a large amount of data to function properly, with many successful real-world applications relying on simulations to train them~\citep{bellemare2020autonomous,degrave2022magnetic}. Some efforts have been made to make agent-based simulations for public policy problems~\citep{zheng2020ai,vardavas2019rand}. But it is unclear whether such simulations have enough accuracy or fidelity to prove useful for training RL policies in simulation and applying them to real public policy or law tasks.

\boxexample{
\begin{example}
[\textbf{Allocating public health resources during a pandemic}] Infectious disease outbreaks can cause severe resource allocation issues. Effective distribution of tests and vaccines are of paramount importance for curtailing the spread of the disease.  But conditions in this setting drift rapidly: one population could quickly become vaccinated and less vulnerable, or a new variant could quickly make another population more susceptible to infection. Could the incorporation of external information like trends from other regions help adjust to drift quickly?
\end{example}
\textsc{Formulation.} Recently, this problem has been formulated as a multi-armed bandit~\citep{chugg2021evaluation} whereby testing resources were allocated to different neighborhoods---each constituting an arm---rewarded by the number of COVID-19 positive individuals identified.
Testing of incoming populations at the border has been similarly formulated as an RL problem, leading to more efficient detection of potentially infectious persons~\citep{bastani2021efficient}. 
However, we might augment such strategies by incorporating the output of disease models into the predicted reward. 
Traditional bandit strategies such as Thompson sampling \citep{russo2017tutorial} or UCB sampling \citep{auer2002using} predict subsequent rewards $\hat{r}_t$ as some function of the history of past rewards $\hat{r}_t = f(\mathcal{H}_t)$. But a disease model $M(\hat{\theta})$ can forecast the effects current case rates on future case rates, and take into account possible effects of new strains and policy changes. In this way, predicted rewards can be anticipative instead of reactive, i.e., $\hat{r}_t=f(M(\hat{\theta}))$ where parameters $\hat{\theta}$ are learned from $\mathcal{H}_t$. Further, if $\hat{\theta}_i$ are the model parameters of location $i$ then, in this context, the explore-exploit tradeoff becomes a tradeoff between approximating the true parameters $\theta_i$, versus allocating resources the most at-risk areas, i.e., that $i$ was where $M(\hat{\theta}_i)$ makes the most severe predictions (i.e., the health equity objective noted above). 
}

\boxexample{
\begin{example}
[\textbf{Inspecting food safety at the FDA}]     The U.S. imports nearly 15 million yearly shipments of food.\footnote{\url{https://www.fda.gov/media/120585/download}} 
    The Food Safety Modernization Act (FSMA) provides for inspections of  imported food products and American-based food production facilities to ensure compliance with FDA standards.\footnote{\url{https://www.fda.gov/media/78021/download}} 
    As acts are amended, however, compliance standards change as do the inspection requirements (see the Tester-Hagan amendment to the FSMA~\citep{boys2015food}, for instance) It may be required to report new information, or to report old information in new ways. This causes distribution shift, and may cause problems for algorithms trained on older inspections data. 
\end{example}
     
     \textsc{Formulation.} Let the arms $\X_t$ be the set of facilities or shipments in year $t$, where each observation containing relevant features and operational details. 
     The hidden label $r_{it}$ for the facility or shipments indicates non-compliance (whether as a binary variable or some measure of the severity). There are various kinds of distribution shift of which we should be wary. The first is simply a change in the relationship of features and labels caused by factors such as price changes, or updated standards in facilities themselves. Models trained to recognize certain attributes as being indicative of unsafe food production might begin to under perform. The second originates from
     statutory amendments or rule changes,  which occur periodically. These can can pose a new kind of distribution shift that we might call \emph{feature shift}: The feature space changes abruptly. A model trained on $\X_t$ might not even be well-defined on $\X_{t+1}$ (e.g., due to a new reporting standard, or the elimination of features). This raises the question of how to take advantage of the information in $\X_t$ for a model which must make predictions on $\X_{t+1}$. Feature engineering and dimensionality reductions are options, but lack theoretical guarantees in the SDM setting.   
}

\subsection{Learning with and Identifying Corrupted Labels}
In many law and public policy settings there is often a ``human in the loop,'' responsible for providing the labels or collecting the reward (e.g., auditors, inspectors, judges). This can inject variance and subjectivity into the information received by the algorithm, which at best can make learning difficult and at worst can make an algorithmic approach infeasible (see Section~\ref{sec:harm}).
This leads to a general warning, illustrated by the example below: when we receive information from heterogeneous and subjective sources, algorithmic design must pay particular attention to the signal-to-noise ratio. In some public sector settings, noise may swamp the signal and make naive machine learning a poor vehicle for solving the problem \citep{kahneman2021noise}. 

There has been a substantial amount of work on learning in the presence of corrupted data and noisy labels, mostly in the classification setting  work on active learning with corrupted labels \citep{naghshvar2012noisy, yan2016active, chen2021corruption,younesian2021qactor}, but more recently in regression as well \citep{chen2020online}. Such work, however, generally ignores the structure of the noisy labels, attempting to only learn the true distribution. In policy settings, however, the structure of the corrupted data may be of interest. We may want to learn the noisy distribution, develop algorithms to determine when a noisy label is likely to occur, or differentiate multiple sources of noise from one another. 

\boxexample{
\begin{example}[\textbf{Prioritizing inspections.}]
Recent work has suggested that machine learning could aid in targeting health and safety inspections~\citep{athey2017beyond,glaeser2016crowdsourcing,johnson2020improving}. Like auditing at other agencies, this process could in theory be improved by SDM. But inspections can be highly stochastic. Recent evidence shows that two health inspectors could give drastically different scores to the same restaurant~\citep{ho2017equity}. Occupational safety and health, mining, nursing home, and nuclear safety inspections each exhibit such inspector variability \citep{ho2017does}. As a result, any algorithms using these noisy (or incorrect) labels may allocate more resources to areas where inspectors are most strict, perpetuating biases and undermining regulatory goals. Until SDM algorithms incorporate mechanisms to identify and correct for such noisy labels, such algorithms should not be deployed blindly in inspection contexts.
\end{example}

\textsc{Formulation.} 
Each timestep, we have inspectors $1,\dots,m$ and facilities with features $x_1,\dots,x_n$. While each facility $x_i$ admits a true reward $r_i$, we receive a noisy signal $r_i+\xi_j$ when inspector $j$ inspects the facility. We attempt to learn both the true rewards and the noise distribution for each inspector. 
Note that if we are able to select which inspector inspects which facility, this adds a layer of complexity on top of the classic bandit model, as there are now two actions to choose but we receive a single confounded reward. We might incorporate this by introducing more terms into our objective function: 
\[\min_{S_1,\dots,S_t}\reg(t) + \alpha\sum_{j=1}^m \kl{\hat{D}_j}{D_j},\]
where $D_j$ is the true noise distribution for inspector $j$, $\hat{D}_j$ is our approximate distribution, and $\alpha$ is some normalizing parameter. Here $\kl{D_1}{D_2}$ is the KL divergence between distributions $D_1$ and $D_2$. 
}

\subsection{Feedback Loops} 
\label{sec:feedback_loops}
 SDMs must be designed to avoid runaway feedback loops---a miscalibrated algorithm may focus only on the areas (or companies, individuals, etc.), at the expense of learning more about others. \citet{ensign2018runaway}, for instance, show  that predictive policing algorithms could repeatedly focus on the same neighbourhoods, even when underlying crime rates are random. Similar effects have also been observed in health care and recommender systems~\citep{sinha2016deconvolving,adam2020hidden,jiang2019degenerate}. 
Recent work has also proposed methods to tackle feedback loops in the context of the bank loan problem~\citep{pacchiano2021neural}. 
Open research questions remain. The most basic is perhaps the question of how to formalize unwanted feedback loops. \citet{jiang2019degenerate} define feedback loops in recommender systems as the condition 
$\limsup_{t\to\infty} \norm{\mu_t-\mu_0}=\infty,$
where $\mu_t$ is the user's interest at time $t$. In other words, the user's interest diverges infinitely from their original preferences. We might change this somewhat for generic multi-armed bandit settings and write  $\limsup_{t\to\infty}\kl{D_t^a}{D_0^a}>T,$
for some finite threshold $T$, where $D_t^a$ is the award distribution of arm $a$ at time $t$, and, as before, $\kl{D_1}{D_2}$ is the KL Divergence between distributions $D_1$ and $D_2$. 
However, such a definition assumes that the feedback loop affects the reward distribution. This might not be the case, and we give an example of another definition in the box below. Moreover, should one adopt this definition, more needs to be done for the problem formulation. In particular, such a definition is coupled with the goal of maximizing reward may encourage feedback loops if they increase the mean of the reward distribution. 

Even after properly defining feedback loops, it's important to ask how we can best (i) detect feedback loops, and (ii) intervene to avoid them.  How can we determine when \emph{not} to use an SDM system if there is a risk for such a feedback loop? Conversely, when does formalization of an existing \emph{ad hoc} SDM system lead to a reduction of existing feedback loops?

\boxexample{
\begin{example}[\textbf{Checking Environmental Compliance at the EPA}] 
\label{ex:epa}

The federal {{Environmental Protection Agency (EPA)}} and state EPAs must decide which facilities to inspect in order to ensure compliance with environmental laws such as the Clean Water Act, Clean Air Act, Safe Drinking Water Act, Toxic Substances Control Act, among others~\citep{10.1145/3442188.3445873}. 
Most of these inspections involve physical visitations which are time intensive and costly. As a result, only a handful of investigation and testing resources are allocated to various facilities. 
The same risk exists in this setting as in predictive policing---algorithms might fall into a local maximum whereby they repeatedly allocate inspections to the same facilities.  
\end{example}

\textsc{Formulation.} Each facility is an arm with features $X_t$ indicating prior enforcement history, region, sensor metrics, calculated risk based on satellite imagery \citep{youssef2011flash,chugg2021enhancing}, and $r_{it}$ is either the amount of unpermitted pollution per facility (regression), or whether the facility was in noncompliance with specific regulations (classification). 
Let $S_t$ be the set of facilities inspected at time $t$. We might identify a positive feedback loop if there exist facilities $i,j$
such that $j$ pollutes more often than $i$, and yet $i$ is visited at the expense of $j$. This might be formalized as the three conditions
\[\sum_{\tau>0} \ind(r_{jt}-r_{it}>0)=\infty,\] 
and 
\[\frac{1}{T}\sum_{\tau=1}^T\ind(i\in S_\tau)\xrightarrow{T\to\infty}1,\quad \frac{1}{T}\sum_{\tau=1}^T \ind(j\in S_\tau)\xrightarrow{T\to\infty}0. \] 
where $\ind(\cdot)$ is an indicator. Note that  $S_t$ is a function of the particular SDM algorithm. We can thereby compare the propensity of different algorithms to generate feedback loops.  
}

\subsection{Rational and Causal Decision-Making}
Law and public policy often impose restrictions on permissible solutions to decision-making problems. Subject matter expertise may also rule out certain solutions. For instance,  in the health setting,
\citet{caruana2015intelligible} discovered that a model trained to predict pneumonia risk learned to associate an asthmatic condition with a \emph{lower} risk of death. This relationship did exist in the training data, but only because asthmatics diagnosed with pneumonia were treated quickly and thoroughly due to their underlying medical condition. The model hence learned precisely the wrong relationship. While this was only a static classification setting, were resources to be dynamically allocated based on such a model, the resulting distribution would be sub-optimal.   

Legal requirements can pose restrictions on solutions. Algorithmic fairness can be conceived of as the area most concerned with the consideration and impacts on protected attributes. But the law may also have requirements for the rationale. An oft-cited requirement of administrative decision-making is that it not be ``arbitrary and capricious.'' 
While the applicability to algorithms adopted by government agencies is open to debate,\footnote{See~\citet{engstrom2020algorithmic,coglianese2019transparency}.} one interpretation is that decisions should be causally sound and not based on spurious correlations, even if such correlations give high predictive accuracy.

There have been several research avenues dedicated to such issues in SDM algorithms. There has been recent work on incorporating prior constraints in reinforcement learning~\citep{roy2021direct}, constrained linear bandits \citep{amani2019linear}, and causal inference in bandits~\citep{lee2018structural,ortega2014generalized,lattimore2016causal,dimakopoulou2019balanced,zhan2021policy,lu2021causal}. 
However, there are some difficulties in applying such work to the kinds of problems we've outlined here. 
For one, causal bandits are focused on learning the effect of interventions on causal graphs. Each arm corresponds to assigning specific values to a set of nodes, after which we witness the effect on the outcome variable. This approach is useful for traditional questions of causal inference, such as determining the effect of school choice on educational outcomes, for instance. It may be less applicable, however, to the kinds of resource-constraint problems we've identified here. For these problems, we're often interested in the causal connection between an observation's features and its label. Whether this problem can be connected to the causal bandits literature in some way is an open question. Meanwhile, the constrained reinforcement learning approach of \citet{roy2021direct} relies on constrained MDPs \citep{altman1999constrained} and uses cost functions to model constraints. The policy space is then restricted to those policies that do not exceed in the cost function in expectation. While promising, meeting constraints in expectation only may be unsuitable for sensitive questions of law and public policy. Finally, constrained bandits restrict the action space to a ``safe set'' of arms. In our applications however, it is rare that constraints take the the form of overt restrictions on what can be sampled. Rather, they pertain to \emph{why} certain arms are sampled, or which arms can be sampled in conjunction with which others (e.g., enforcing limits on how many facilities from a given geographic can be audited). Moreover, the approaches outlined above, insofar as they are applicable to public policy, must also be adapted to deal with the difficulties of previous sections  (distribution shifts, unknown causal structures, small data, multiple objectives, etc.). 

\boxexample{
\begin{example}[\textbf{Identifying Claims at the SSA}]
    To reduce processing times, the U.S. Social Security Administration (SSA) has developed mechanisms for quickly identifying disability claims likely to be approved~\citep{engstrom2020government}. The system makes its prediction based on information such as medical history and treatment protocols. 
    \end{example}

\textsc{Formulation.} 
While there may be some \emph{a priori} identifiable constraints -- e.g., perhaps that identification of disability claims should not depend on gender -- others must be learned over time. Specifically, administrative law may require ascertaining whether the algorithm  (i) has relied on impermissible factors, (ii) has incorporated required factors (e.g., those that the law mandates be taken into account), and (iii) has reached a reasoned decision. These raise challenging technical problems. Suppose we have a model $f_{\hat{\theta}}$ used to predict whether a disability claim $x$ will be approved. At each round $t$, the model receives a new claim and determines whether to flag it for review or not. If it is reviewed, a human will identify the true label and the model will be updated. At round $t$, we discover that $f_{\hat{\theta}}$ has been relying on a correlation that it shouldn't. How do we modify $f_{\hat{\theta}}$ to adopt new constraints? Simply removing the confounds $C$ from the feature space may be insufficient, as other features may be correlated with $C$. One possibility involves \emph{adversarial selection}, originally proposed for learning deconfounded lexicons in NLP \citep{pryzant2018deconfounded}. The idea is that a decision rule for $f_{\hat{\theta}}$ should be unrelated to the confounds $C$, thus have trouble predicting $C$ given $x$. We can implement this using a gradient reversal layer \citep{ganin2015unsupervised}. More specifically, we train a neural network to learn an encoding $e(x)$ of the observation $x$ whose loss is $\mathcal{L}_y-\mathcal{L}_C$, where $\mathcal{L}_y$ is the loss of $e(x)$ predicting the label $y$, and $\mathcal{L}_C$ the loss of $e(x)$ predicting values of $C$. Thus, we learn an encoding which is correlated with $y$ but not with $C$. This approach has also been used to try and enforce fairness constraints in neural networks \citep{raff2018gradient}. As this is a sequential decision making system, these selection mechanisms may need to be introduced into the sampling algorithm itself, as in the case of balanced bandits~\citep{dimakopoulou2019balanced}.
}

\subsection{Validation}
\label{sec:validation}

Even when all of the above challenges are tackled, researchers working on SDMs for public policy face another not-entirely-solved hurdle: validation of the SDM algorithm.

Validation for SDM algorithms remains an ongoing and unsolved challenge. To be confident that a particular SDM method will function as expected, it would ideally provide confidence intervals on validation performance, theoretical guarantees, and some notion of robustness to distribution shifts. Researchers will likely need to do this \emph{a priori} with logged data to be sure of an algorithm's utility.

The right method for measuring and providing confidence intervals on the performance of SDM algorithms is still debated~\citep{henderson2018deep,agarwal2021deep}.
Off-policy evaluation on restrospective data is also an ongoing research challenge~\citep{dudik2012sample,narita2021debiased}.
And much work has pointed to the consistently underpowered evaluations of machine learning algorithms~\citep{colas2018many,agarwal2021deep,card2020little}.
Finally, providing statistical guarantees on algorithm performance often using relying on assumptions of linearity, with only some work proving guarantees for non-linear estimators or agnostic of estimators~\citep{dong2021provable}.

Policy problems and associated data do not conform to formats that are ideal for validation. Data is often high-dimensional and non-linear~\citep{henderson2022maximize}. Though sometimes it is randomly collected, it may also come from data that could be highly off-policy. And more often than not, there is little data to create benchmarks with large statistical power. As such, it may be necessary to incorporate evaluation in the SDM algorithm itself to ensure sufficient exploration to assess algorithm performance in an iterative fashion. Recent work explores such challenges and may prove fruitful in tackling the validation difficulties in law and public policy ~\citep{yao2021power,zanette2021design,chandak2021universal}.

The ability to evaluate an algorithm in a robust fashion has implications for ensuring compliance with legal standards -- a unique challenge for public policy and law deployments. As noted earlier, the law may require population estimates. The 2018 Office of Management and Budget (OMB) guidelines, for example, recommended that compliance with regulations required misreporting estimates to be ``statistically valid'' (unbiased estimates of the mean) and have ``$\pm 3\%$ or better margin of error at the 95\% confidence level  for  the  improper  payment  percentage  estimate.''~\citep{tas2018improper,omb2018requirements}.
And new legal requirements for algorithmic robustness are being introduced (see, e.g., a survey of legislation and regulation efforts in \citep{publicsectorai}), that may introduce requirements on validation procedures.

Last, validation may be easier with simpler approaches \cite{rudin2019stop}, which may inform the complexity of modeling. While standard bandits, for instance, do not require \emph{modeling} features or context, contextual bandits or reinforcement learning approaches typically do. Such models may require more investigation to validate (e.g., to assess model sensitivity, calibration, tuning). In part for that reason, one public health jurisdiction adopted a simpler bandit approach for allocating testing \cite{chugg2021evaluation}. 

\boxexample{
\begin{example}[\textbf{Wildfire Prevention Targeting at DOI}]
    The U.S. Department of the Interior (DOI) is responsible for implementing wildfire management policies. A 2019 Executive Order tasks DOI with implementing more active management of forests to improve wildfire outcomes~\citep{us2018promoting}. In particular, DOI takes several wildfire fuel treatment actions to reduce the magnitude of forest fires, including prescribed fires, removing brush, etc. DOI utilizes decision support systems to prioritize and track treatment of fuels, in addition to implementing a database to validate predicted outcomes of those treatments when a forest fire hits the treated fuel~\citep{wildfirepolicymemo}. 
    \end{example}

\textsc{Formulation.} 
    Various SDM algorithms have been applied to wildfire management policy-making and others have reviewed a number of these applications, including reinforcement learning methods~\citep{jain2020review}. \citet{lauer2017spatial}, for example, propose a reinforcement learning for learning optimal strategies for forest management to reduce wildfire risk. Consider an RL problem, where an agent at timestep $x$ is given the option to clear brush or conduct a prescribed burn from one of $A$ regions. It is difficult to reward the agent based on true wildfire reduction, so the agent is trained in simulation with a custom reward as prescribed by \citet{lauer2017spatial} to reduce wildfire risk. There is limited data to evaluate the effectiveness of policy based on recorded wildfire intensities after treatments, provided by the Fuels Treatment Effectiveness Monitoring program~\citep{wildfirepolicymemo}. Key open questions are how researchers can evaluate a program program based on  limited off-policy data without  overfitting.
}

\section{Assessing and mitigating social harms}
\label{sec:harm}

Ensuring the safety and reliability of machine learning systems used in government is of paramount importance due to the sensitivity of information being collected and the importance of the decisions being made. If automated decision making systems are relied on at scale, small errors can be magnified in their effect size~\cite{brauneis2018algorithmic}. Algorithmic failure modes have been amply demonstrated, from bias and disparate impact~\citep{barocas2016big, buolamwini2018gender, coston2021leveraging, khalil2020investigating, mehrabi2021survey, obermeyer2019dissecting}, lack of transparency~\citep{coglianese2019transparency,janssen2016challenges, rudin2019stop}, erosion of privacy~\citep{berman2018government, tucker2018privacy}, and misplaced or non-existent accountability~\cite{coglianese2021contracting, engstrom2020algorithmic, kroll2015accountable}. Such issues have spurred much discussion on the role and safety measures of AI in government~\cite{rubenstein2021acquiring, calo2017artificial, mulligan2019procurement,lehr2017playing} in addition to the legality of employing these tools~\cite{coglianese2016regulating, engstrom2020algorithmic}. We refrain from restating these known and serious potential harms and refer readers to extensive treatments on these topics~\citep{kearns2019ethical, barocas2017fairness}, but we note that  addressing such harms will be critical in considering any SDM deployment. From a validity perspective~\citep{coston2022validity}, our work highlights the technical challenges required for successful public sector deployment. The advertising fallacy can lead to harms if these challenges aren't catalogued and faced head on, and the motivations and potential benefits set forth in Section~\ref{sec:motivation} may be irrelevant if harms are not addressed.

We do emphasize that SDM algorithms carry with them distinct risks of social harm, beyond those well-documented in non-sequential machine learning. SDM algorithms like reinforcement learning may learn to exploit reward functions in unpredictable ways. \citet{carroll2022estimating}, for example, found that there are situations where an algorithm can learn to induce preference shifts in its users to achieve its goals. In the context of a recommendation system this may mean radicalizing a user to get more clicks, as opposed to simply finding content that the user might like. 
The influences that these effects can have on policymakers and line officers at agencies, may be in conflict with core legal principles~\citep{engstrom2020algorithmic}. As illustrated in Section \ref{sec:feedback_loops}, the danger of feedback loops in SDM algorithms is substantial. The danger arises precisely because, as witnessed by the many applications we've discussed, an SDM algorithm is selecting its own training data. This results in statistical bias which, if unaccounted for, can lead the algorithm to begin self-reinforcing its own priors. 
Poor initial validation can lead to model overconfidence, and thus over-reliance on the predictions by policymakers. This problem can be exacerbated by small datasets which may cause overfitting and a tendency to read erroneous signals from noise. High quality, ground truth datasets remain essential to overcome cold start problems with SDMs. 

While the examples we discuss here may illustrate potential benefits of formulating existing processes as SDMs, 
the same techniques might be applied for more nefarious purposes. Efficient algorithms might, for instance, encourage government overreach, or be used by autocratic regimes. They might also falsely convince policymakers that better algorithms are a one-size-fits-all solution, but they cannot substitute for good governance and strong oversight.

Finally, when policies can be updated, bad actors might learn to exploit algorithmic failure modes. For example, they might inject poisoned data~\citep{ma2018data,liu2019data} to steer the algorithm away from them or to manipulate outcomes.
Consider for example the case of audit selection. If a set of entities wished to exploit the algorithm and avoid audits, entities with similar features could pool their resources together and vigorously fight any audits for their group. If the algorithm saw enough examples of audits that yielded no successful enforcement outcomes, it might de-prioritize the entire group in future rounds. This could then lead to a feedback loop where the group is never audited again.

Several implications follow from these harms.  First, the deployment of an SDM algorithm should  be accompanied by a cost-benefit analysis, rolled out incrementally, and robustly stress-tested before large-scale deployment. 
Machine learning developers must work with public sector partners to identify the right mechanisms for oversight. 

Second, shifting from off-the-shelf one-time procurements to a continuous oversight is essential for SDM systems.
Such auditing mechanisms can potentially be encoded directly into the SDM process (e.g., inducing exploration to yield proper estimates of validity at each timestep).
We note that further research is needed into auditing and monitoring~\citep{citron2007technological, raji}. For example, to date, relying on FOIA requests has been insufficient for full transparency~\citep{fink2018opening}.
In short, the adoption of SDMs in the public sector heighten the need for mechanisms of review, evaluation, and oversight.

\section{Conclusion}

Our work has reviewed the potential and challenges for SDM in the public sector.
SDM algorithms have the promise to improve law and public policy, making the public sector more efficient, transparent, and accountable. But the research challenges posed by the public sector are considerable to render such approaches policy-compliant, trustworthy, and effective. We hope this review will inspire much more work to address these challenges at the intersection of law and computer science and identify the potential social gains and limitations of SDMs in law and policy.

\bibliographystyle{plainnat}
 \balance
\bibliography{sample-base.bib}

\begin{thebibliography}{159}
\providecommand{\natexlab}[1]{#1}
\providecommand{\url}[1]{\texttt{#1}}
\expandafter\ifx\csname urlstyle\endcsname\relax
  \providecommand{\doi}[1]{doi: #1}\else
  \providecommand{\doi}{doi: \begingroup \urlstyle{rm}\Url}\fi

\bibitem[{Ada Lovelace Institute, AI Now Institute and Open Government
  Partnership}(2021)]{publicsectorai}
{Ada Lovelace Institute, AI Now Institute and Open Government Partnership}.
\newblock Algorithmic accountability for the public sector.
\newblock 2021.

\bibitem[Adam et~al.(2020)Adam, Chang, Haibe-Kains, and
  Goldenberg]{adam2020hidden}
George~Alexandru Adam, Chun-Hao~Kingsley Chang, Benjamin Haibe-Kains, and Anna
  Goldenberg.
\newblock Hidden risks of machine learning applied to healthcare: Unintended
  feedback loops between models and future data causing model degradation.
\newblock In \emph{Machine Learning for Healthcare Conference}, pages 710--731.
  PMLR, 2020.

\bibitem[Agarwal et~al.(2021)Agarwal, Schwarzer, Castro, Courville, and
  Bellemare]{agarwal2021deep}
Rishabh Agarwal, Max Schwarzer, Pablo~Samuel Castro, Aaron~C Courville, and
  Marc Bellemare.
\newblock Deep reinforcement learning at the edge of the statistical precipice.
\newblock \emph{Advances in Neural Information Processing Systems}, 34, 2021.

\bibitem[Alami et~al.(2017)Alami, Maillard, and F{\'e}raud]{alami2017memory}
R{\'e}da Alami, Odalric Maillard, and Raphael F{\'e}raud.
\newblock Memory bandits: a bayesian approach for the switching bandit problem.
\newblock In \emph{NIPS 2017-31st Conference on Neural Information Processing
  Systems}, 2017.

\bibitem[Altman(1999)]{altman1999constrained}
Eitan Altman.
\newblock \emph{Constrained Markov decision processes: stochastic modeling}.
\newblock Routledge, 1999.

\bibitem[Amani et~al.(2019)Amani, Alizadeh, and Thrampoulidis]{amani2019linear}
Sanae Amani, Mahnoosh Alizadeh, and Christos Thrampoulidis.
\newblock Linear stochastic bandits under safety constraints.
\newblock \emph{Advances in Neural Information Processing Systems}, 32, 2019.

\bibitem[Amarasinghe et~al.(2021)Amarasinghe, Rodolfa, Lamba, and
  Ghani]{amarasinghe2021explainable}
Kasun Amarasinghe, Kit Rodolfa, Hemank Lamba, and Rayid Ghani.
\newblock Explainable machine learning for public policy: Use cases, gaps, and
  research directions, 2021.

\bibitem[Andriotis and Papakonstantinou(2021)]{andriotis2021deep}
CP~Andriotis and KG~Papakonstantinou.
\newblock Deep reinforcement learning driven inspection and maintenance
  planning under incomplete information and constraints.
\newblock \emph{Reliability Engineering \& System Safety}, 212:\penalty0
  107551, 2021.

\bibitem[Antos et~al.(2010)Antos, Grover, and Szepesv{\'a}ri]{antos2010active}
Andr{\'a}s Antos, Varun Grover, and Csaba Szepesv{\'a}ri.
\newblock Active learning in heteroscedastic noise.
\newblock \emph{Theoretical Computer Science}, 411\penalty0 (29-30):\penalty0
  2712--2728, 2010.

\bibitem[Ash et~al.(2021)Ash, Galletta, and Giommoni]{ash2021machine}
Elliott Ash, Sergio Galletta, and Tommaso Giommoni.
\newblock A machine learning approach to analyze and support anti-corruption
  policy.
\newblock 2021.

\bibitem[Athey(2017)]{athey2017beyond}
Susan Athey.
\newblock Beyond prediction: Using big data for policy problems.
\newblock \emph{Science}, 355\penalty0 (6324):\penalty0 483--485, 2017.

\bibitem[Auer(2002)]{auer2002using}
Peter Auer.
\newblock Using confidence bounds for exploitation-exploration trade-offs.
\newblock \emph{Journal of Machine Learning Research}, 3\penalty0
  (Nov):\penalty0 397--422, 2002.

\bibitem[Barocas and Selbst(2016)]{barocas2016big}
Solon Barocas and Andrew~D Selbst.
\newblock Big data's disparate impact.
\newblock \emph{Calif. L. Rev.}, 104:\penalty0 671, 2016.

\bibitem[Barocas et~al.()Barocas, Hardt, and Narayanan]{barocas2017fairness}
Solon Barocas, Moritz Hardt, and Arvind Narayanan.
\newblock Fairness in machine learning.

\bibitem[Bastani and Bayati(2020)]{bastani2020online}
Hamsa Bastani and Mohsen Bayati.
\newblock Online decision making with high-dimensional covariates.
\newblock \emph{Operations Research}, 68\penalty0 (1):\penalty0 276--294, 2020.

\bibitem[Bastani et~al.(2021)Bastani, Drakopoulos, Gupta, Vlachogiannis,
  Hadjicristodoulou, Lagiou, Magiorkinis, Paraskevis, and
  Tsiodras]{bastani2021efficient}
Hamsa Bastani, Kimon Drakopoulos, Vishal Gupta, Jon Vlachogiannis, Christos
  Hadjicristodoulou, Pagona Lagiou, Gkikas Magiorkinis, Dimitrios Paraskevis,
  and Sotirios Tsiodras.
\newblock Efficient and targeted {COVID-19} border testing via reinforcement
  learning, 2021.

\bibitem[Beam and Kohane(2018)]{beam2018big}
Andrew~L Beam and Isaac~S Kohane.
\newblock Big data and machine learning in health care.
\newblock \emph{Jama}, 319\penalty0 (13):\penalty0 1317--1318, 2018.

\bibitem[Bellemare et~al.(2020)Bellemare, Candido, Castro, Gong, Machado,
  Moitra, Ponda, and Wang]{bellemare2020autonomous}
Marc~G Bellemare, Salvatore Candido, Pablo~Samuel Castro, Jun Gong, Marlos~C
  Machado, Subhodeep Moitra, Sameera~S Ponda, and Ziyu Wang.
\newblock Autonomous navigation of stratospheric balloons using reinforcement
  learning.
\newblock \emph{Nature}, 588\penalty0 (7836):\penalty0 77--82, 2020.

\bibitem[Bellman(1957)]{bellman1957markovian}
Richard Bellman.
\newblock A markovian decision process.
\newblock \emph{Journal of mathematics and mechanics}, pages 679--684, 1957.

\bibitem[Benami et~al.(2021)Benami, Whitaker, La, Lin, Anderson, and
  Ho]{10.1145/3442188.3445873}
Elinor Benami, Reid Whitaker, Vincent La, Hongjin Lin, Brandon~R. Anderson, and
  Daniel~E. Ho.
\newblock The distributive effects of risk prediction in environmental
  compliance: Algorithmic design, environmental justice, and public policy.
\newblock In \emph{Proceedings of the 2021 ACM Conference on Fairness,
  Accountability, and Transparency}, FAccT '21, page 90–105, New York, NY,
  USA, 2021. Association for Computing Machinery.
\newblock ISBN 9781450383097.
\newblock \doi{10.1145/3442188.3445873}.
\newblock URL \url{https://doi.org/10.1145/3442188.3445873}.

\bibitem[Berman(2018)]{berman2018government}
Emily Berman.
\newblock A government of laws and not of machines.
\newblock \emph{Bul rev.}, 98:\penalty0 1277, 2018.

\bibitem[Bistritz et~al.(2019)Bistritz, Zhou, Chen, Bambos, and
  Blanchet]{bistritz2019exp3}
Ilai Bistritz, Zhengyuan Zhou, Xi~Chen, Nicholas Bambos, and Jose Blanchet.
\newblock Exp3 learning in adversarial bandits with delayed feedback.
\newblock \emph{Advances in neural information processing systems}, 2019.

\bibitem[Boys et~al.(2015)Boys, Ollinger, and Geyer]{boys2015food}
Kathryn~A Boys, Michael Ollinger, and Leon~L Geyer.
\newblock The food safety modernization act: implications for us small scale
  farms.
\newblock \emph{American Journal of Law \& Medicine}, 41\penalty0
  (2-3):\penalty0 395--405, 2015.

\bibitem[Brauneis and Goodman(2018)]{brauneis2018algorithmic}
Robert Brauneis and Ellen~P Goodman.
\newblock Algorithmic transparency for the smart city.
\newblock \emph{Yale JL \& Tech.}, 20:\penalty0 103, 2018.

\bibitem[Buolamwini and Gebru(2018)]{buolamwini2018gender}
Joy Buolamwini and Timnit Gebru.
\newblock Gender shades: Intersectional accuracy disparities in commercial
  gender classification.
\newblock In \emph{Conference on fairness, accountability and transparency},
  pages 77--91. PMLR, 2018.

\bibitem[Calo(2017)]{calo2017artificial}
Ryan Calo.
\newblock Artificial intelligence policy: a primer and roadmap.
\newblock \emph{UCDL Rev.}, 51:\penalty0 399, 2017.

\bibitem[Card et~al.(2020)Card, Henderson, Khandelwal, Jia, Mahowald, and
  Jurafsky]{card2020little}
Dallas Card, Peter Henderson, Urvashi Khandelwal, Robin Jia, Kyle Mahowald, and
  Dan Jurafsky.
\newblock With little power comes great responsibility.
\newblock \emph{arXiv preprint arXiv:2010.06595}, 2020.

\bibitem[Carpentier et~al.(2011)Carpentier, Lazaric, Ghavamzadeh, Munos, and
  Auer]{carpentier2011upper}
Alexandra Carpentier, Alessandro Lazaric, Mohammad Ghavamzadeh, R{\'e}mi Munos,
  and Peter Auer.
\newblock Upper-confidence-bound algorithms for active learning in multi-armed
  bandits.
\newblock In \emph{International Conference on Algorithmic Learning Theory},
  pages 189--203. Springer, 2011.

\bibitem[Carroll et~al.(2022)Carroll, Dragan, Russell, and
  Hadfield-Menell]{carroll2022estimating}
Micah~D Carroll, Anca Dragan, Stuart Russell, and Dylan Hadfield-Menell.
\newblock Estimating and penalizing induced preference shifts in recommender
  systems.
\newblock In \emph{International Conference on Machine Learning}, pages
  2686--2708. PMLR, 2022.

\bibitem[Caruana et~al.(2015)Caruana, Lou, Gehrke, Koch, Sturm, and
  Elhadad]{caruana2015intelligible}
Rich Caruana, Yin Lou, Johannes Gehrke, Paul Koch, Marc Sturm, and Noemie
  Elhadad.
\newblock Intelligible models for healthcare: Predicting pneumonia risk and
  hospital 30-day readmission.
\newblock In \emph{Proceedings of the 21th ACM SIGKDD international conference
  on knowledge discovery and data mining}, pages 1721--1730, 2015.

\bibitem[Cassandra(1998)]{cassandra1998survey}
Anthony~R Cassandra.
\newblock A survey of {POMDP} applications.
\newblock In \emph{Working notes of AAAI 1998 fall symposium on planning with
  partially observable Markov decision processes}, volume 1724, 1998.

\bibitem[Cavenaghi et~al.(2021)Cavenaghi, Sottocornola, Stella, and
  Zanker]{cavenaghi2021non}
Emanuele Cavenaghi, Gabriele Sottocornola, Fabio Stella, and Markus Zanker.
\newblock Non stationary multi-armed bandit: Empirical evaluation of a new
  concept drift-aware algorithm.
\newblock \emph{Entropy}, 23\penalty0 (3):\penalty0 380, 2021.

\bibitem[Chandak et~al.(2021)Chandak, Niekum, da~Silva, Learned-Miller,
  Brunskill, and Thomas]{chandak2021universal}
Yash Chandak, Scott Niekum, Bruno da~Silva, Erik Learned-Miller, Emma
  Brunskill, and Philip~S Thomas.
\newblock Universal off-policy evaluation.
\newblock \emph{Advances in Neural Information Processing Systems}, 34, 2021.

\bibitem[Chen et~al.(2020)Chen, Koehler, Moitra, and Yau]{chen2020online}
Sitan Chen, Frederic Koehler, Ankur Moitra, and Morris Yau.
\newblock Online and distribution-free robustness: Regression and contextual
  bandits with huber contamination.
\newblock \emph{arXiv preprint arXiv:2010.04157}, 2020.

\bibitem[Chen et~al.(2013)Chen, Wang, and Yuan]{chen2013combinatorial}
Wei Chen, Yajun Wang, and Yang Yuan.
\newblock Combinatorial multi-armed bandit: General framework and applications.
\newblock In \emph{International Conference on Machine Learning}, pages
  151--159. PMLR, 2013.

\bibitem[Chen et~al.(2021)Chen, Du, and Jamieson]{chen2021corruption}
Yifang Chen, Simon~S Du, and Kevin Jamieson.
\newblock Corruption robust active learning.
\newblock \emph{arXiv preprint arXiv:2106.11220}, 2021.

\bibitem[Chhatwal et~al.(2017)Chhatwal, Huber-Fliflet, Keeling, Zhang, and
  Zhao]{chhatwal2017empirical}
Rishi Chhatwal, Nathaniel Huber-Fliflet, Robert Keeling, Jianping Zhang, and
  Haozhen Zhao.
\newblock Empirical evaluations of active learning strategies in legal document
  review.
\newblock In \emph{2017 IEEE International Conference on Big Data (Big Data)},
  pages 1428--1437. IEEE, 2017.

\bibitem[Chugg and Ho(2021)]{chugg2021reconciling}
Ben Chugg and Daniel~E Ho.
\newblock Reconciling risk allocation and prevalence estimation in public
  health using batched bandits.
\newblock \emph{arXiv preprint arXiv:2110.13306}, 2021.

\bibitem[Chugg et~al.(2021{\natexlab{a}})Chugg, Anderson, Eicher, Lee, and
  Ho]{chugg2021enhancing}
Ben Chugg, Brandon Anderson, Seiji Eicher, Sandy Lee, and Daniel~E Ho.
\newblock Enhancing environmental enforcement with near real-time monitoring:
  Likelihood-based detection of structural expansion of intensive livestock
  farms.
\newblock \emph{International Journal of Applied Earth Observation and
  Geoinformation}, 103:\penalty0 102463, 2021{\natexlab{a}}.

\bibitem[Chugg et~al.(2021{\natexlab{b}})Chugg, Lu, Ouyang, Anderson, Ha,
  D’Agostino, Sujeer, Rudman, Garcia, and Ho]{chugg2021evaluation}
Ben Chugg, Lisa Lu, Derek Ouyang, Benjamin Anderson, Raymond Ha, Alexis
  D’Agostino, Anandi Sujeer, Sarah~L Rudman, Analilia Garcia, and Daniel~E
  Ho.
\newblock Evaluation of allocation schemes of {COVID}-19 testing resources in a
  community-based door-to-door testing program.
\newblock In \emph{JAMA Health Forum}, volume~2, pages e212260--e212260.
  American Medical Association, 2021{\natexlab{b}}.

\bibitem[Citovsky et~al.(2021)Citovsky, DeSalvo, Gentile, Karydas, Rajagopalan,
  Rostamizadeh, and Kumar]{citovsky2021batch}
Gui Citovsky, Giulia DeSalvo, Claudio Gentile, Lazaros Karydas, Anand
  Rajagopalan, Afshin Rostamizadeh, and Sanjiv Kumar.
\newblock Batch active learning at scale.
\newblock \emph{arXiv preprint arXiv:2107.14263}, 2021.

\bibitem[Citron(2007)]{citron2007technological}
Danielle~Keats Citron.
\newblock Technological due process.
\newblock \emph{Wash. UL Rev.}, 85:\penalty0 1249, 2007.

\bibitem[Coglianese and Lampmann(2021)]{coglianese2021contracting}
Cary Coglianese and Erik Lampmann.
\newblock Contracting for algorithmic accountability.
\newblock \emph{Administrative Law Review Accord}, 6:\penalty0 175, 2021.

\bibitem[Coglianese and Lehr(2016)]{coglianese2016regulating}
Cary Coglianese and David Lehr.
\newblock Regulating by robot: Administrative decision making in the
  machine-learning era.
\newblock \emph{Geo. LJ}, 105:\penalty0 1147, 2016.

\bibitem[Coglianese and Lehr(2019)]{coglianese2019transparency}
Cary Coglianese and David Lehr.
\newblock Transparency and algorithmic governance.
\newblock \emph{Admin. L. Rev.}, 71:\penalty0 1, 2019.

\bibitem[Colas et~al.(2018)Colas, Sigaud, and Oudeyer]{colas2018many}
C{\'e}dric Colas, Olivier Sigaud, and Pierre-Yves Oudeyer.
\newblock How many random seeds? statistical power analysis in deep
  reinforcement learning experiments.
\newblock \emph{arXiv preprint arXiv:1806.08295}, 2018.

\bibitem[Cormack and Grossman(2015)]{cormack2015autonomy}
Gordon~V Cormack and Maura~R Grossman.
\newblock Autonomy and reliability of continuous active learning for
  technology-assisted review.
\newblock \emph{arXiv preprint arXiv:1504.06868}, 2015.

\bibitem[Coston et~al.(2021)Coston, Guha, Ouyang, Lu, Chouldechova, and
  Ho]{coston2021leveraging}
Amanda Coston, Neel Guha, Derek Ouyang, Lisa Lu, Alexandra Chouldechova, and
  Daniel~E Ho.
\newblock Leveraging administrative data for bias audits: Assessing disparate
  coverage with mobility data for covid-19 policy.
\newblock In \emph{Proceedings of the 2021 ACM Conference on Fairness,
  Accountability, and Transparency}, pages 173--184, 2021.

\bibitem[Coston et~al.(2022)Coston, Kawakami, Zhu, Holstein, and
  Heidari]{coston2022validity}
Amanda Coston, Anna Kawakami, Haiyi Zhu, Ken Holstein, and Hoda Heidari.
\newblock A validity perspective on evaluating the justified use of data-driven
  decision-making algorithms.
\newblock \emph{arXiv preprint arXiv:2206.14983}, 2022.

\bibitem[Degrave et~al.(2022)Degrave, Felici, Buchli, Neunert, Tracey,
  Carpanese, Ewalds, Hafner, Abdolmaleki, de~Las~Casas,
  et~al.]{degrave2022magnetic}
Jonas Degrave, Federico Felici, Jonas Buchli, Michael Neunert, Brendan Tracey,
  Francesco Carpanese, Timo Ewalds, Roland Hafner, Abbas Abdolmaleki, Diego
  de~Las~Casas, et~al.
\newblock Magnetic control of tokamak plasmas through deep reinforcement
  learning.
\newblock \emph{Nature}, 602\penalty0 (7897):\penalty0 414--419, 2022.

\bibitem[Dimakopoulou et~al.(2019)Dimakopoulou, Zhou, Athey, and
  Imbens]{dimakopoulou2019balanced}
Maria Dimakopoulou, Zhengyuan Zhou, Susan Athey, and Guido Imbens.
\newblock Balanced linear contextual bandits.
\newblock \emph{Proceedings of the AAAI Conference on Artificial Intelligence},
  33\penalty0 (01):\penalty0 3445--3453, 2019.

\bibitem[Dong et~al.(2021)Dong, Yang, and Ma]{dong2021provable}
Kefan Dong, Jiaqi Yang, and Tengyu Ma.
\newblock Provable model-based nonlinear bandit and reinforcement learning:
  Shelve optimism, embrace virtual curvature.
\newblock \emph{Advances in Neural Information Processing Systems}, 34, 2021.

\bibitem[Drugan and Nowe(2013)]{drugan2013designing}
Madalina~M Drugan and Ann Nowe.
\newblock Designing multi-objective multi-armed bandits algorithms: A study.
\newblock In \emph{The 2013 International Joint Conference on Neural Networks
  (IJCNN)}, pages 1--8. IEEE, 2013.

\bibitem[Dud{\'\i}k et~al.(2012)Dud{\'\i}k, Erhan, Langford, and
  Li]{dudik2012sample}
Miroslav Dud{\'\i}k, Dumitru Erhan, John Langford, and Lihong Li.
\newblock Sample-efficient nonstationary policy evaluation for contextual
  bandits.
\newblock \emph{arXiv preprint arXiv:1210.4862}, 2012.

\bibitem[Durand et~al.(2018)Durand, Achilleos, Iacovides, Strati, Mitsis, and
  Pineau]{durand2018contextual}
Audrey Durand, Charis Achilleos, Demetris Iacovides, Katerina Strati,
  Georgios~D Mitsis, and Joelle Pineau.
\newblock Contextual bandits for adapting treatment in a mouse model of de novo
  carcinogenesis.
\newblock In \emph{Machine learning for healthcare conference}, pages 67--82.
  PMLR, 2018.

\bibitem[Engstrom and Ho(2020)]{engstrom2020algorithmic}
David~Freeman Engstrom and Daniel~E Ho.
\newblock Algorithmic accountability in the administrative state.
\newblock \emph{Yale J. on Reg.}, 37:\penalty0 800, 2020.

\bibitem[Engstrom et~al.(2020)Engstrom, Ho, Sharkey, and
  Cu{\'e}llar]{engstrom2020government}
David~Freeman Engstrom, Daniel~E Ho, Catherine~M Sharkey, and
  Mariano-Florentino Cu{\'e}llar.
\newblock Government by algorithm: Artificial intelligence in federal
  administrative agencies.
\newblock \emph{NYU School of Law, Public Law Research Paper}, \penalty0
  (20-54), 2020.

\bibitem[Ensign et~al.(2018)Ensign, Friedler, Neville, Scheidegger, and
  Venkatasubramanian]{ensign2018runaway}
Danielle Ensign, Sorelle~A Friedler, Scott Neville, Carlos Scheidegger, and
  Suresh Venkatasubramanian.
\newblock Runaway feedback loops in predictive policing.
\newblock In \emph{Conference on Fairness, Accountability and Transparency},
  pages 160--171. PMLR, 2018.

\bibitem[Erraqabi et~al.(2017)Erraqabi, Lazaric, Valko, Brunskill, and
  Liu]{erraqabi2017trading}
Akram Erraqabi, Alessandro Lazaric, Michal Valko, Emma Brunskill, and Yun-En
  Liu.
\newblock Trading off rewards and errors in multi-armed bandits.
\newblock In \emph{Artificial Intelligence and Statistics}, pages 709--717.
  PMLR, 2017.

\bibitem[Esfandiari et~al.(2021)Esfandiari, Karbasi, Mehrabian, and
  Mirrokni]{esfandiari2021regret}
Hossein Esfandiari, Amin Karbasi, Abbas Mehrabian, and Vahab Mirrokni.
\newblock Regret bounds for batched bandits.
\newblock In \emph{Proceedings of the AAAI Conference on Artificial
  Intelligence}, volume~35, pages 7340--7348, 2021.

\bibitem[Ferraz and Finan(2008)]{ferraz2008exposing}
Claudio Ferraz and Frederico Finan.
\newblock Exposing corrupt politicians: the effects of brazil's publicly
  released audits on electoral outcomes.
\newblock \emph{The Quarterly journal of economics}, 123\penalty0 (2):\penalty0
  703--745, 2008.

\bibitem[Fink(2018)]{fink2018opening}
Katherine Fink.
\newblock Opening the government’s black boxes: freedom of information and
  algorithmic accountability.
\newblock \emph{Information, Communication \& Society}, 21\penalty0
  (10):\penalty0 1453--1471, 2018.

\bibitem[Ganin and Lempitsky(2015)]{ganin2015unsupervised}
Yaroslav Ganin and Victor Lempitsky.
\newblock Unsupervised domain adaptation by backpropagation.
\newblock In \emph{International conference on machine learning}, pages
  1180--1189. PMLR, 2015.

\bibitem[Gao et~al.(2019)Gao, Han, Ren, and Zhou]{gao2019batched}
Zijun Gao, Yanjun Han, Zhimei Ren, and Zhengqing Zhou.
\newblock Batched multi-armed bandits problem.
\newblock \emph{arXiv preprint arXiv:1904.01763}, 2019.

\bibitem[Ghassemi et~al.(2020)Ghassemi, Naumann, Schulam, Beam, Chen, and
  Ranganath]{ghassemi2020review}
Marzyeh Ghassemi, Tristan Naumann, Peter Schulam, Andrew~L Beam, Irene~Y Chen,
  and Rajesh Ranganath.
\newblock A review of challenges and opportunities in machine learning for
  health.
\newblock \emph{AMIA Summits on Translational Science Proceedings},
  2020:\penalty0 191, 2020.

\bibitem[Glaeser et~al.(2016)Glaeser, Hillis, Kominers, and
  Luca]{glaeser2016crowdsourcing}
Edward~L Glaeser, Andrew Hillis, Scott~Duke Kominers, and Michael Luca.
\newblock Crowdsourcing city government: Using tournaments to improve
  inspection accuracy.
\newblock \emph{American Economic Review}, 106\penalty0 (5):\penalty0 114--18,
  2016.

\bibitem[{Government Accountability Office}(2021)]{gao_epa}
{Government Accountability Office}.
\newblock \emph{Clean Water Act: EPA Needs to Better Assess and Disclose
  Quality of Compliance and Enforcement Data}.
\newblock 2021.

\bibitem[Grossman and Cormack(2014)]{grossman2014comments}
Maura~R Grossman and Gordon~V Cormack.
\newblock Comments on “the implications of rule 26 (g) on the use of
  technology-assisted review”.
\newblock \emph{Federal Courts Law Review}, 1, 2014.

\bibitem[Grover et~al.(2018)Grover, Markov, Attia, Jin, Perkins, Cheong, Chen,
  Yang, Harris, Chueh, et~al.]{grover2018best}
Aditya Grover, Todor Markov, Peter Attia, Norman Jin, Nicolas Perkins, Bryan
  Cheong, Michael Chen, Zi~Yang, Stephen Harris, William Chueh, et~al.
\newblock Best arm identification in multi-armed bandits with delayed feedback.
\newblock In \emph{International Conference on Artificial Intelligence and
  Statistics}, pages 833--842. PMLR, 2018.

\bibitem[Guha et~al.(2022)Guha, Henderson, and
  Zambrano]{henderson2022discovery}
Neel Guha, Peter Henderson, and Diego Zambrano.
\newblock Vulnerabilities in discovery tech.
\newblock \emph{Harvard Journal of Law \& Technology}, 2022.

\bibitem[Guo and Schuurmans(2007)]{guo2007discriminative}
Yuhong Guo and Dale Schuurmans.
\newblock Discriminative batch mode active learning.
\newblock In \emph{NIPS}, pages 593--600. Citeseer, 2007.

\bibitem[Henderson et~al.(2018)Henderson, Islam, Bachman, Pineau, Precup, and
  Meger]{henderson2018deep}
Peter Henderson, Riashat Islam, Philip Bachman, Joelle Pineau, Doina Precup,
  and David Meger.
\newblock Deep reinforcement learning that matters.
\newblock In \emph{Proceedings of the AAAI conference on artificial
  intelligence}, volume~32, 2018.

\bibitem[Henderson et~al.(2022)Henderson, Chugg, Anderson, Altenburger, Turk,
  Guyton, Goldin, and Ho]{henderson2022maximize}
Peter Henderson, Ben Chugg, Brandon Anderson, Kristen Altenburger, Alex Turk,
  John Guyton, Jacob Goldin, and Daniel~E. Ho.
\newblock {Integrating Reward Maximization and Population Estimation:
  Sequential Decision-Making for Internal Revenue Service Audit Selection}.
\newblock \emph{arXiv preprint arxiv:2204.11910}, 2022.

\bibitem[Ho(2017{\natexlab{a}})]{ho2017does}
Daniel~E Ho.
\newblock Does peer review work? an experiment of experimentalism.
\newblock \emph{Stanford Law Review}, 69\penalty0 (1):\penalty0 1--120,
  2017{\natexlab{a}}.

\bibitem[Ho(2017{\natexlab{b}})]{ho2017equity}
Daniel~E Ho.
\newblock Equity in the bureaucracy.
\newblock \emph{UC Irvine L. Rev.}, 7:\penalty0 401, 2017{\natexlab{b}}.

\bibitem[Horvitz and Thompson(1952)]{horvitz1952generalization}
Daniel~G Horvitz and Donovan~J Thompson.
\newblock A generalization of sampling without replacement from a finite
  universe.
\newblock \emph{Journal of the American statistical Association}, 47\penalty0
  (260):\penalty0 663--685, 1952.

\bibitem[Huang and Lin(2016)]{huang2016linear}
Kuan-Hao Huang and Hsuan-Tien Lin.
\newblock Linear upper confidence bound algorithm for contextual bandit problem
  with piled rewards.
\newblock In \emph{Pacific-Asia Conference on Knowledge Discovery and Data
  Mining}, pages 143--155. Springer, 2016.

\bibitem[Jain et~al.(2020)Jain, Coogan, Subramanian, Crowley, Taylor, and
  Flannigan]{jain2020review}
Piyush Jain, Sean~CP Coogan, Sriram~Ganapathi Subramanian, Mark Crowley, Steve
  Taylor, and Mike~D Flannigan.
\newblock A review of machine learning applications in wildfire science and
  management.
\newblock \emph{Environmental Reviews}, 28\penalty0 (4):\penalty0 478--505,
  2020.

\bibitem[Janssen and Kuk(2016)]{janssen2016challenges}
Marijn Janssen and George Kuk.
\newblock The challenges and limits of big data algorithms in technocratic
  governance, 2016.

\bibitem[Jiang et~al.(2019)Jiang, Chiappa, Lattimore, Gy{\"o}rgy, and
  Kohli]{jiang2019degenerate}
Ray Jiang, Silvia Chiappa, Tor Lattimore, Andr{\'a}s Gy{\"o}rgy, and Pushmeet
  Kohli.
\newblock Degenerate feedback loops in recommender systems.
\newblock In \emph{Proceedings of the 2019 AAAI/ACM Conference on AI, Ethics,
  and Society}, pages 383--390, 2019.

\bibitem[Johnson et~al.(2020)Johnson, Levine, and Toffel]{johnson2020improving}
Matthew~S Johnson, David~I Levine, and Michael~W Toffel.
\newblock Improving regulatory effectiveness through better targeting: Evidence
  from osha.
\newblock \emph{Harvard Business School Technology \& Operations Mgt. Unit
  Working Paper}, \penalty0 (20-019), 2020.

\bibitem[Jurafsky and Martin(2014)]{jurafsky2014speech}
Dan Jurafsky and James~H Martin.
\newblock Speech and language processing. vol. 3.
\newblock \emph{US: Prentice Hall}, 2014.

\bibitem[Kabir et~al.(2014)Kabir, Sadiq, and Tesfamariam]{kabir2014review}
Golam Kabir, Rehan Sadiq, and Solomon Tesfamariam.
\newblock A review of multi-criteria decision-making methods for infrastructure
  management.
\newblock \emph{Structure and infrastructure engineering}, 10\penalty0
  (9):\penalty0 1176--1210, 2014.

\bibitem[Kahneman et~al.(2021)Kahneman, Sibony, and
  Sunstein]{kahneman2021noise}
Daniel Kahneman, Olivier Sibony, and Cass~R Sunstein.
\newblock \emph{Noise: A flaw in human judgment}.
\newblock Little, Brown, 2021.

\bibitem[Kaminski and Malgieri(2020)]{kaminski2020algorithmic}
Margot~E Kaminski and Gianclaudio Malgieri.
\newblock Algorithmic impact assessments under the gdpr: producing
  multi-layered explanations.
\newblock \emph{International Data Privacy Law}, pages 19--28, 2020.

\bibitem[Karpatne et~al.(2018)Karpatne, Ebert-Uphoff, Ravela, Babaie, and
  Kumar]{karpatne2018machine}
Anuj Karpatne, Imme Ebert-Uphoff, Sai Ravela, Hassan~Ali Babaie, and Vipin
  Kumar.
\newblock Machine learning for the geosciences: Challenges and opportunities.
\newblock \emph{IEEE Transactions on Knowledge and Data Engineering},
  31\penalty0 (8):\penalty0 1544--1554, 2018.

\bibitem[Kearns and Roth(2019)]{kearns2019ethical}
Michael Kearns and Aaron Roth.
\newblock \emph{The ethical algorithm: The science of socially aware algorithm
  design}.
\newblock Oxford University Press, 2019.

\bibitem[Khalil et~al.(2020)Khalil, Ahmed, Khattak, and
  Al-Qirim]{khalil2020investigating}
Ashraf Khalil, Soha~Glal Ahmed, Asad~Masood Khattak, and Nabeel Al-Qirim.
\newblock Investigating bias in facial analysis systems: A systematic review.
\newblock \emph{IEEE Access}, 8:\penalty0 130751--130761, 2020.

\bibitem[Krass et~al.(2021)Krass, Henderson, Mello, Studdert, and
  Ho]{krass2021us}
Mark Krass, Peter Henderson, Michelle~M Mello, David~M Studdert, and Daniel~E
  Ho.
\newblock How us law will evaluate artificial intelligence for covid-19.
\newblock \emph{bmj}, 372, 2021.

\bibitem[Krawczyk and Cano(2019)]{krawczyk2019adaptive}
Bartosz Krawczyk and Alberto Cano.
\newblock Adaptive ensemble active learning for drifting data stream mining.
\newblock In \emph{IJCAI}, pages 2763--2771, 2019.

\bibitem[Kroll(2015)]{kroll2015accountable}
Joshua~Alexander Kroll.
\newblock \emph{Accountable algorithms}.
\newblock PhD thesis, Princeton University, 2015.

\bibitem[Lansdell et~al.(2019)Lansdell, Triantafillou, and
  Kording]{lansdell2019rarely}
Benjamin Lansdell, Sofia Triantafillou, and Konrad Kording.
\newblock Rarely-switching linear bandits: optimization of causal effects for
  the real world.
\newblock \emph{arXiv preprint arXiv:1905.13121}, 2019.

\bibitem[Lattimore et~al.(2016)Lattimore, Lattimore, and
  Reid]{lattimore2016causal}
Finnian Lattimore, Tor Lattimore, and Mark~D Reid.
\newblock Causal bandits: learning good interventions via causal inference.
\newblock In \emph{Proceedings of the 30th International Conference on Neural
  Information Processing Systems}, pages 1189--1197, 2016.

\bibitem[Lauer et~al.(2017)Lauer, Montgomery, and Dietterich]{lauer2017spatial}
Christopher~J Lauer, Claire~A Montgomery, and Thomas~G Dietterich.
\newblock Spatial interactions and optimal forest management on a
  fire-threatened landscape.
\newblock \emph{Forest Policy and Economics}, 83:\penalty0 107--120, 2017.

\bibitem[Lee and Bareinboim(2018)]{lee2018structural}
Sanghack Lee and Elias Bareinboim.
\newblock Structural causal bandits: where to intervene?
\newblock \emph{Advances in Neural Information Processing Systems 31}, 31,
  2018.

\bibitem[Lehr and Ohm(2017)]{lehr2017playing}
David Lehr and Paul Ohm.
\newblock Playing with the data: what legal scholars should learn about machine
  learning.
\newblock \emph{UCDL Rev.}, 51:\penalty0 653, 2017.

\bibitem[Li and Kanoulas(2020)]{li2020stop}
Dan Li and Evangelos Kanoulas.
\newblock When to stop reviewing in technology-assisted reviews: Sampling from
  an adaptive distribution to estimate residual relevant documents.
\newblock \emph{ACM Transactions on Information Systems (TOIS)}, 38\penalty0
  (4):\penalty0 1--36, 2020.

\bibitem[Li et~al.(2010)Li, Chu, Langford, and Schapire]{li2010contextual}
Lihong Li, Wei Chu, John Langford, and Robert~E Schapire.
\newblock A contextual-bandit approach to personalized news article
  recommendation.
\newblock In \emph{Proceedings of the 19th international conference on World
  wide web}, pages 661--670, 2010.

\bibitem[Liu and Shroff(2019)]{liu2019data}
Fang Liu and Ness Shroff.
\newblock Data poisoning attacks on stochastic bandits.
\newblock In \emph{International Conference on Machine Learning}, pages
  4042--4050. PMLR, 2019.

\bibitem[Liu et~al.(2014)Liu, Mandel, Brunskill, and Popovic]{liu2014trading}
Yun-En Liu, Travis Mandel, Emma Brunskill, and Zoran Popovic.
\newblock Trading off scientific knowledge and user learning with multi-armed
  bandits.
\newblock In \emph{EDM}, pages 161--168, 2014.

\bibitem[Lohr(2019)]{lohr2019sampling}
Sharon~L Lohr.
\newblock \emph{Sampling: design and analysis}.
\newblock Chapman and Hall/CRC, 2019.

\bibitem[Lu et~al.(2021)Lu, Meisami, and Tewari]{lu2021causal}
Yangyi Lu, Amirhossein Meisami, and Ambuj Tewari.
\newblock Causal bandits with unknown graph structure.
\newblock \emph{arXiv preprint arXiv:2106.02988}, 2021.

\bibitem[Ma et~al.(2018)Ma, Jun, Li, and Zhu]{ma2018data}
Yuzhe Ma, Kwang-Sung Jun, Lihong Li, and Xiaojin Zhu.
\newblock Data poisoning attacks in contextual bandits.
\newblock In \emph{International Conference on Decision and Game Theory for
  Security}, pages 186--204. Springer, 2018.

\bibitem[Madanat(1993)]{madanat1993optimal}
Samer Madanat.
\newblock Optimal infrastructure management decisions under uncertainty.
\newblock \emph{Transportation Research Part C: Emerging Technologies},
  1\penalty0 (1):\penalty0 77--88, 1993.

\bibitem[McKelvey and MacDonald(2019)]{mckelvey2019artificial}
Fenwick McKelvey and Margaret MacDonald.
\newblock Artificial intelligence policy innovations at the canadian federal
  government.
\newblock \emph{Canadian Journal of Communication}, 44\penalty0 (2):\penalty0
  PP43--PP50, 2019.

\bibitem[Mehrabi et~al.(2021)Mehrabi, Morstatter, Saxena, Lerman, and
  Galstyan]{mehrabi2021survey}
Ninareh Mehrabi, Fred Morstatter, Nripsuta Saxena, Kristina Lerman, and Aram
  Galstyan.
\newblock A survey on bias and fairness in machine learning.
\newblock \emph{ACM Computing Surveys (CSUR)}, 54\penalty0 (6):\penalty0 1--35,
  2021.

\bibitem[Mellor and Shapiro(2013)]{mellor2013thompson}
Joseph Mellor and Jonathan Shapiro.
\newblock Thompson sampling in switching environments with bayesian online
  change detection.
\newblock In \emph{Artificial Intelligence and Statistics}, pages 442--450.
  PMLR, 2013.

\bibitem[Mersereau et~al.(2009)Mersereau, Rusmevichientong, and
  Tsitsiklis]{mersereau2009structured}
Adam~J Mersereau, Paat Rusmevichientong, and John~N Tsitsiklis.
\newblock A structured multiarmed bandit problem and the greedy policy.
\newblock \emph{IEEE Transactions on Automatic Control}, 54\penalty0
  (12):\penalty0 2787--2802, 2009.

\bibitem[Metcalf et~al.(2021)Metcalf, Moss, Watkins, Singh, and
  Elish]{metcalf2021algorithmic}
Jacob Metcalf, Emanuel Moss, Elizabeth~Anne Watkins, Ranjit Singh, and
  Madeleine~Clare Elish.
\newblock Algorithmic impact assessments and accountability: The
  co-construction of impacts.
\newblock In \emph{Proceedings of the 2021 ACM Conference on Fairness,
  Accountability, and Transparency}, pages 735--746, 2021.

\bibitem[Morantz(2009)]{morantz2009has}
Alison~D Morantz.
\newblock Has devolution injured american workers? state and federal
  enforcement of construction safety.
\newblock \emph{The Journal of Law, Economics, \& Organization}, 25\penalty0
  (1):\penalty0 183--210, 2009.

\bibitem[Moss et~al.(2020)Moss, Watkins, Metcalf, and Elish]{moss2020governing}
Emanuel Moss, Elizabeth~Anne Watkins, Jacob Metcalf, and Madeleine~Clare Elish.
\newblock Governing with algorithmic impact assessments: six observations.
\newblock \emph{Available at SSRN 3584818}, 2020.

\bibitem[Mulligan and Bamberger(2019)]{mulligan2019procurement}
Deirdre~K Mulligan and Kenneth~A Bamberger.
\newblock Procurement as policy: Administrative process for machine learning.
\newblock \emph{Berkeley Tech. LJ}, 34:\penalty0 773, 2019.

\bibitem[Naghshvar et~al.(2012)Naghshvar, Javidi, and
  Chaudhuri]{naghshvar2012noisy}
Mohammad Naghshvar, Tara Javidi, and Kamalika Chaudhuri.
\newblock Noisy bayesian active learning.
\newblock In \emph{2012 50th Annual Allerton Conference on Communication,
  Control, and Computing (Allerton)}, pages 1626--1633. IEEE, 2012.

\bibitem[Narita et~al.(2021)Narita, Yasui, and Yata]{narita2021debiased}
Yusuke Narita, Shota Yasui, and Kohei Yata.
\newblock Debiased off-policy evaluation for recommendation systems.
\newblock In \emph{Fifteenth ACM Conference on Recommender Systems}, pages
  372--379, 2021.

\bibitem[Nie et~al.(2018)Nie, Tian, Taylor, and Zou]{nie2018adaptively}
Xinkun Nie, Xiaoying Tian, Jonathan Taylor, and James Zou.
\newblock Why adaptively collected data have negative bias and how to correct
  for it.
\newblock In \emph{International Conference on Artificial Intelligence and
  Statistics}, pages 1261--1269. PMLR, 2018.

\bibitem[Nogas et~al.(2021)Nogas, Li, Yanez, Modiri, Deliu, Prystawski, Villar,
  Rafferty, and Williams]{nogas2021algorithms}
Jacob Nogas, Tong Li, Fernando~J Yanez, Arghavan Modiri, Nina Deliu, Ben
  Prystawski, Sofia~S Villar, Anna Rafferty, and Joseph~J Williams.
\newblock Algorithms for adaptive experiments that trade-off statistical
  analysis with reward: Combining uniform random assignment and reward
  maximization.
\newblock \emph{arXiv preprint arXiv:2112.08507}, 2021.

\bibitem[Obermeyer et~al.(2019)Obermeyer, Powers, Vogeli, and
  Mullainathan]{obermeyer2019dissecting}
Ziad Obermeyer, Brian Powers, Christine Vogeli, and Sendhil Mullainathan.
\newblock Dissecting racial bias in an algorithm used to manage the health of
  populations.
\newblock \emph{Science}, 366\penalty0 (6464):\penalty0 447--453, 2019.

\bibitem[{Office of Management and Budget}(2018)]{omb2018requirements}
{Office of Management and Budget}.
\newblock Requirements for payment integrity improvement.
\newblock
  \url{https://www.whitehouse.gov/wp-content/uploads/2018/06/M-18-20.pdf},
  2018.
\newblock Executive Office of the President. Online; Accessed Jan 10, 2022.

\bibitem[Ortega and Braun(2014)]{ortega2014generalized}
Pedro~A Ortega and Daniel~A Braun.
\newblock Generalized thompson sampling for sequential decision-making and
  causal inference.
\newblock \emph{Complex Adaptive Systems Modeling}, 2\penalty0 (1):\penalty0
  1--23, 2014.

\bibitem[Pacchiano et~al.(2021)Pacchiano, Singh, Chou, Berg, and
  Foerster]{pacchiano2021neural}
Aldo Pacchiano, Shaun Singh, Edward Chou, Alex Berg, and Jakob Foerster.
\newblock Neural pseudo-label optimism for the bank loan problem.
\newblock \emph{Advances in Neural Information Processing Systems}, 34, 2021.

\bibitem[Papernot et~al.(2016)Papernot, McDaniel, Sinha, and
  Wellman]{papernot2016towards}
Nicolas Papernot, Patrick McDaniel, Arunesh Sinha, and Michael Wellman.
\newblock Towards the science of security and privacy in machine learning.
\newblock \emph{arXiv preprint arXiv:1611.03814}, 2016.

\bibitem[Perchet et~al.(2016)Perchet, Rigollet, Chassang, and
  Snowberg]{perchet2016batched}
Vianney Perchet, Philippe Rigollet, Sylvain Chassang, and Erik Snowberg.
\newblock Batched bandit problems.
\newblock \emph{The Annals of Statistics}, 44\penalty0 (2):\penalty0 660--681,
  2016.

\bibitem[Pryzant et~al.(2018)Pryzant, Shen, Jurafsky, and
  Wagner]{pryzant2018deconfounded}
Reid Pryzant, Kelly Shen, Dan Jurafsky, and Stefan Wagner.
\newblock Deconfounded lexicon induction for interpretable social science.
\newblock In \emph{Proceedings of the 2018 Conference of the North American
  Chapter of the Association for Computational Linguistics: Human Language
  Technologies, Volume 1 (Long Papers)}, pages 1615--1625, 2018.

\bibitem[Puterman(1994)]{Puterman94}
Martin~L. Puterman.
\newblock \emph{Markov Decision Processes---Discrete Stochastic Dynamic
  Programming}.
\newblock John Wiley \& Sons, Inc., New York, NY, 1994.

\bibitem[Raff and Sylvester(2018)]{raff2018gradient}
Edward Raff and Jared Sylvester.
\newblock Gradient reversal against discrimination.
\newblock \emph{arXiv preprint arXiv:1807.00392}, 2018.

\bibitem[Raji et~al.(2022)Raji, Xu, Honigsberg, and Ho]{raji}
Inioluwa~Deborah Raji, Peggy Xu, Colleen Honigsberg, and Daniel Ho.
\newblock Outsider oversight: Designing a third party audit ecosystem for ai
  governance.
\newblock In \emph{Proceedings of the 2022 AAAI/ACM Conference on AI, Ethics,
  and Society}, AIES '22, page 557–571, New York, NY, USA, 2022. Association
  for Computing Machinery.
\newblock ISBN 9781450392471.
\newblock \doi{10.1145/3514094.3534181}.
\newblock URL \url{https://doi.org/10.1145/3514094.3534181}.

\bibitem[Reyes et~al.(2009)Reyes, Spaan, and Sucar]{reyes2009intelligent}
Alberto Reyes, Matthijs~TJ Spaan, and L~Enrique Sucar.
\newblock An intelligent assistant for power plants based on factored {MDP}s.
\newblock In \emph{2009 15th International Conference on Intelligent System
  Applications to Power Systems}, pages 1--6. IEEE, 2009.

\bibitem[Roy et~al.(2021)Roy, Girgis, Romoff, Bacon, and Pal]{roy2021direct}
Julien Roy, Roger Girgis, Joshua Romoff, Pierre-Luc Bacon, and Christopher Pal.
\newblock Direct behavior specification via constrained reinforcement learning.
\newblock \emph{arXiv preprint arXiv:2112.12228}, 2021.

\bibitem[Rubenstein(2021)]{rubenstein2021acquiring}
David~S Rubenstein.
\newblock Acquiring ethical ai.
\newblock \emph{Fla. L. Rev.}, 73:\penalty0 747, 2021.

\bibitem[Rudin(2019)]{rudin2019stop}
Cynthia Rudin.
\newblock Stop explaining black box machine learning models for high stakes
  decisions and use interpretable models instead.
\newblock \emph{Nature Machine Intelligence}, 1\penalty0 (5):\penalty0
  206--215, 2019.

\bibitem[Rupert(2020)]{wildfirepolicymemo}
Jeffrey Rupert.
\newblock Doi wildland fire program policy memorandum no. 2020-004.
\newblock 2020.

\bibitem[Russac et~al.(2019)Russac, Vernade, and Capp{\'e}]{russac2019weighted}
Yoan Russac, Claire Vernade, and Olivier Capp{\'e}.
\newblock Weighted linear bandits for non-stationary environments.
\newblock In \emph{NeurIPS 2019-33rd Conference on Neural Information
  Processing Systems}, 2019.

\bibitem[Russo et~al.(2017)Russo, Van~Roy, Kazerouni, Osband, and
  Wen]{russo2017tutorial}
Daniel Russo, Benjamin Van~Roy, Abbas Kazerouni, Ian Osband, and Zheng Wen.
\newblock A tutorial on thompson sampling.
\newblock \emph{arXiv preprint arXiv:1707.02038}, 2017.

\bibitem[Sarin and Summers(2019)]{sarin2019shrinking}
Natasha Sarin and Lawrence~H Summers.
\newblock Shrinking the tax gap: approaches and revenue potential.
\newblock Technical report, National Bureau of Economic Research, 2019.

\bibitem[Scholz et~al.(2014)Scholz, Levihn, Isbell, and
  Wingate]{scholz2014physics}
Jonathan Scholz, Martin Levihn, Charles Isbell, and David Wingate.
\newblock A physics-based model prior for object-oriented {MDP}s.
\newblock In \emph{International Conference on Machine Learning}, pages
  1089--1097. PMLR, 2014.

\bibitem[Settles(2009)]{settles2009active}
Burr Settles.
\newblock Active learning literature survey, 2009.

\bibitem[Shann and Seuken(2014)]{shann2014adaptive}
Mike Shann and Sven Seuken.
\newblock Adaptive home heating under weather and price uncertainty using {GP}s
  and {MDP}s.
\newblock In \emph{Proceedings of the 2014 international conference on
  Autonomous agents and multi-agent systems}, pages 821--828, 2014.

\bibitem[Shin et~al.(2019)Shin, Ramdas, and Rinaldo]{shin2019sample}
Jaehyeok Shin, Aaditya Ramdas, and Alessandro Rinaldo.
\newblock Are sample means in multi-armed bandits positively or negatively
  biased?
\newblock \emph{Advances in Neural Information Processing Systems},
  32:\penalty0 7102--7111, 2019.

\bibitem[Shin et~al.(2021)Shin, Ramdas, and Rinaldo]{shin2021bias}
Jaehyeok Shin, Aaditya Ramdas, and Alessandro Rinaldo.
\newblock On the bias, risk, and consistency of sample means in multi-armed
  bandits.
\newblock \emph{SIAM Journal on Mathematics of Data Science}, 3\penalty0
  (4):\penalty0 1278--1300, 2021.

\bibitem[Shyam et~al.(2019)Shyam, Ja{\'s}kowski, and Gomez]{shyam2019model}
Pranav Shyam, Wojciech Ja{\'s}kowski, and Faustino Gomez.
\newblock Model-based active exploration.
\newblock In \emph{International conference on machine learning}, pages
  5779--5788. PMLR, 2019.

\bibitem[Sinha et~al.(2016)Sinha, Gleich, and Ramani]{sinha2016deconvolving}
Ayan Sinha, David~F Gleich, and Karthik Ramani.
\newblock Deconvolving feedback loops in recommender systems.
\newblock \emph{Advances in neural information processing systems},
  29:\penalty0 3243--3251, 2016.

\bibitem[Soemers et~al.(2018)Soemers, Brys, Driessens, Winands, and
  Now{\'e}]{soemers2018adapting}
Dennis Soemers, Tim Brys, Kurt Driessens, Mark Winands, and Ann Now{\'e}.
\newblock Adapting to concept drift in credit card transaction data streams
  using contextual bandits and decision trees.
\newblock In \emph{Proceedings of the AAAI Conference on Artificial
  Intelligence}, volume~32, 2018.

\bibitem[Tang et~al.(2014)Tang, Jiang, Li, and Li]{tang2014ensemble}
Liang Tang, Yexi Jiang, Lei Li, and Tao Li.
\newblock Ensemble contextual bandits for personalized recommendation.
\newblock In \emph{Proceedings of the 8th ACM Conference on Recommender
  Systems}, pages 73--80, 2014.

\bibitem[Tarbouriech and Lazaric(2019)]{tarbouriech2019active}
Jean Tarbouriech and Alessandro Lazaric.
\newblock Active exploration in markov decision processes.
\newblock In \emph{The 22nd International Conference on Artificial Intelligence
  and Statistics}, pages 974--982. PMLR, 2019.

\bibitem[{Taxpayer Advocate Service}(2018)]{tas2018improper}
{Taxpayer Advocate Service}.
\newblock Improper earned income tax credit payments: Measures the {IRS} takes
  to reduce improper earned income tax credit payments are not sufficiently
  proactive and may unnecessarily burden taxpayers.
\newblock
  \url{https://www.taxpayeradvocate.irs.gov/wp-content/uploads/2020/07/ARC18_Volume1_MSP_06_ImproperEarnedIncome.pdf},
  2018.
\newblock 2018 Annual Report to Congress — Volume One. Online; Accessed Jan
  10, 2022.

\bibitem[Thrun et~al.(1991)Thrun, M{\"o}ller, and Linden]{thrun1991active}
Sebastian~B Thrun, Knut M{\"o}ller, and A~Linden.
\newblock Active exploration in dynamic environments.
\newblock In \emph{NIPS}, pages 531--538, 1991.

\bibitem[Tucker(2018)]{tucker2018privacy}
Catherine Tucker.
\newblock Privacy, algorithms, and artificial intelligence.
\newblock In \emph{The economics of artificial intelligence: An agenda}, pages
  423--437. University of Chicago Press, 2018.

\bibitem[{US Executive Office of the President}(2018)]{us2018promoting}
{US Executive Office of the President}.
\newblock {Promoting Active Management of America’s Forests, Rangelands, and
  Other Federal Lands To Improve Conditions and Reduce Wildfire Risk (Executive
  Order 13855)}.
\newblock 2018.

\bibitem[Vardavas et~al.(2019)Vardavas, Katkar, Parker, Aliyev, Graf, and
  Kumar]{vardavas2019rand}
Raffaele Vardavas, Pavan Katkar, Andrew~M Parker, GR~Aliyev, Marlon Graf, and
  KB~Kumar.
\newblock Rand’s interdisciplinary behavioral and social science agent-based
  model of income tax evasion.
\newblock 2019.

\bibitem[Vernade et~al.(2020)Vernade, Carpentier, Lattimore, Zappella, Ermis,
  and Brueckner]{vernade2020linear}
Claire Vernade, Alexandra Carpentier, Tor Lattimore, Giovanni Zappella, Beyza
  Ermis, and Michael Brueckner.
\newblock Linear bandits with stochastic delayed feedback.
\newblock In \emph{International Conference on Machine Learning}, pages
  9712--9721. PMLR, 2020.

\bibitem[Warofka(2018)]{warofka2018independent}
Alex Warofka.
\newblock An independent assessment of the human rights impact of facebook in
  myanmar.
\newblock \emph{Facebook Newsroom, November}, 5, 2018.

\bibitem[Yan et~al.(2016)Yan, Chaudhuri, and Javidi]{yan2016active}
Songbai Yan, Kamalika Chaudhuri, and Tara Javidi.
\newblock Active learning from imperfect labelers.
\newblock \emph{Advances in Neural Information Processing Systems},
  29:\penalty0 2128--2136, 2016.

\bibitem[Yao et~al.(2021)Yao, Brunskill, Pan, Murphy, and
  Doshi-Velez]{yao2021power}
Jiayu Yao, Emma Brunskill, Weiwei Pan, Susan Murphy, and Finale Doshi-Velez.
\newblock Power constrained bandits.
\newblock In \emph{Machine Learning for Healthcare Conference}, pages 209--259.
  PMLR, 2021.

\bibitem[Younesian et~al.(2021)Younesian, Zhao, Ghiassi, Birke, and
  Chen]{younesian2021qactor}
Taraneh Younesian, Zilong Zhao, Amirmasoud Ghiassi, Robert Birke, and Lydia~Y
  Chen.
\newblock Qactor: Active learning on noisy labels.
\newblock In \emph{Asian Conference on Machine Learning}, pages 548--563. PMLR,
  2021.

\bibitem[Youssef et~al.(2011)Youssef, Pradhan, and Hassan]{youssef2011flash}
Ahmed~M Youssef, Biswajeet Pradhan, and Abdallah~Mohamed Hassan.
\newblock Flash flood risk estimation along the st. katherine road, southern
  sinai, egypt using gis based morphometry and satellite imagery.
\newblock \emph{Environmental Earth Sciences}, 62\penalty0 (3):\penalty0
  611--623, 2011.

\bibitem[Zanette et~al.(2021)Zanette, Dong, Lee, and
  Brunskill]{zanette2021design}
Andrea Zanette, Kefan Dong, Jonathan Lee, and Emma Brunskill.
\newblock Design of experiments for stochastic contextual linear bandits.
\newblock \emph{Advances in Neural Information Processing Systems}, 34, 2021.

\bibitem[Zhan et~al.(2021)Zhan, Ren, Athey, and Zhou]{zhan2021policy}
Ruohan Zhan, Zhimei Ren, Susan Athey, and Zhengyuan Zhou.
\newblock Policy learning with adaptively collected data.
\newblock \emph{arXiv preprint arXiv:2105.02344}, 2021.

\bibitem[Zhao et~al.(2020)Zhao, Zhang, Jiang, and Zhou]{zhao2020simple}
Peng Zhao, Lijun Zhang, Yuan Jiang, and Zhi-Hua Zhou.
\newblock A simple approach for non-stationary linear bandits.
\newblock In \emph{International Conference on Artificial Intelligence and
  Statistics}, pages 746--755. PMLR, 2020.

\bibitem[Zheng et~al.(2020)Zheng, Trott, Srinivasa, Naik, Gruesbeck, Parkes,
  and Socher]{zheng2020ai}
Stephan Zheng, Alexander Trott, Sunil Srinivasa, Nikhil Naik, Melvin Gruesbeck,
  David~C Parkes, and Richard Socher.
\newblock The ai economist: Improving equality and productivity with ai-driven
  tax policies.
\newblock \emph{arXiv preprint arXiv:2004.13332}, 2020.

\end{thebibliography}

\end{document}